\documentclass[10pt,journal,letter]{IEEEtran}
\UseRawInputEncoding
\usepackage{multirow}

\usepackage{graphicx}
\usepackage{mathtools}
\usepackage{commath}
\usepackage{amsmath}
\usepackage{graphicx}
\usepackage{algorithm}
\usepackage{algpseudocode}
\usepackage{algorithmicx}
\usepackage{epstopdf}
\usepackage{multirow}
\usepackage{array}
\usepackage{pgfplots}
\begin{document}
\title{Auditory Model based Phase-Aware Bayesian Spectral Amplitude Estimator for Single-Channel Speech Enhancement}

\author{Suman~Samui,~\IEEEmembership{Student Member,~IEEE,}
        Indrajit~Chakrabarti,~\IEEEmembership{Member,~IEEE,}
        and~Soumya~K.~Ghosh,~\IEEEmembership{Member,~IEEE}
\thanks{S. Samui is with Advanced Technology Development Centre, Indian Institute of Technology, Kharagpur (e-mail: samuisuman@gmail.com), I. Chakrabarti is with Department of Electronics and Electrical Communication Engineering, IIT Kharagpur (e-mail: indrajit@ece.iitkgp.ac.in), and S. K. Ghosh is with Department of Computer Science and Engineering, IIT Kharagpur, West Bengal-721302, India (e-mail: skg@iitkgp.ac.in).}}

\maketitle

\begin{abstract}
Bayesian estimation of short-time spectral amplitude (STSA) is one of the most predominant approaches for the enhancement of the noise corrupted speech. The performance of these estimators are usually significantly improved when any perceptually relevant cost function is considered. On the other hand, the recent progress in phase-based speech signal processing have shown that the phase-only enhancement based on spectral phase estimation methods can also provide joint improvement in the perceived speech quality and intelligibility, even in low SNR conditions. In this paper, to take advantage of both the perceptually motivated cost function involving STSAs of estimated and true clean speech and utilizing the prior spectral phase information, we have derived a phase-aware Bayesian STSA estimator. The parameters of the cost function are chosen based on the the characteristics of the human auditory system, namely, the dynamic compressive non-linearity of the cochlea, the perceived loudness theory and the simultaneous masking properties of ear. This type of parameter selection scheme results in more noise reduction while limiting the speech distortion. The derived STSA estimator is optimal in the MMSE sense if the prior phase information is available. In practice, however, typically only an estimate of the clean speech phase can be obtained via employing  different types of spectral phase estimation techniques which have been developed throughout the last few years. In a blind setup, we have evaluated the proposed Bayesian STSA estimator with different types of standard phase estimation methods available in the literature. Experimental results have shown that the proposed estimator can achieve substantial improvement in performance than the traditional phase-blind approaches as well as the other existing phase-aware Bayesian estimators such as those based on the $\beta$-order minimum mean-square error (MMSE) of the STSA in terms of various objective measures.
\end{abstract}


\section{Introduction}
\IEEEPARstart{D}{ue} to the wide range of applications, single-channel speech enhancement (SCSE) has attracted significant amount of research attention for more than last three decades. The main objective of SCSE is to operate on a monaural noisy observation (speech samples) to reduce the corrupting noise component (improving quality) while preserving the speech intelligibility as much as possible. Signal processing based SCSE research has a long-standing history and it has led to the development of numerous proposals in spectral \cite{parchami2016recent}\cite{loizou2013}, modulation \cite{Paliwal2012}, and temporal \cite{yegnanarayana1999speech}\cite{samui2016two}\cite{tavares2016speech} domains.  However, none of the well-known techniques of speech enhancement was deemed to be promising in improving the speech intelligibility relative to unprocessed corrupted speech \cite{hu2007}, particularly at low SNR conditions. Recently,  the deep neural network (DNN) based data-driven speech enhancement systems have shown remarkable potential in improving the speech quality and intelligibility simultaneously, but the generalization of these DNN models in various dimensions (SNRs, noise types and speakers) is still considered as an open problem and it requires more research attention \cite{sie2017}.
\begin{figure}
\centering
\includegraphics[scale=0.35]{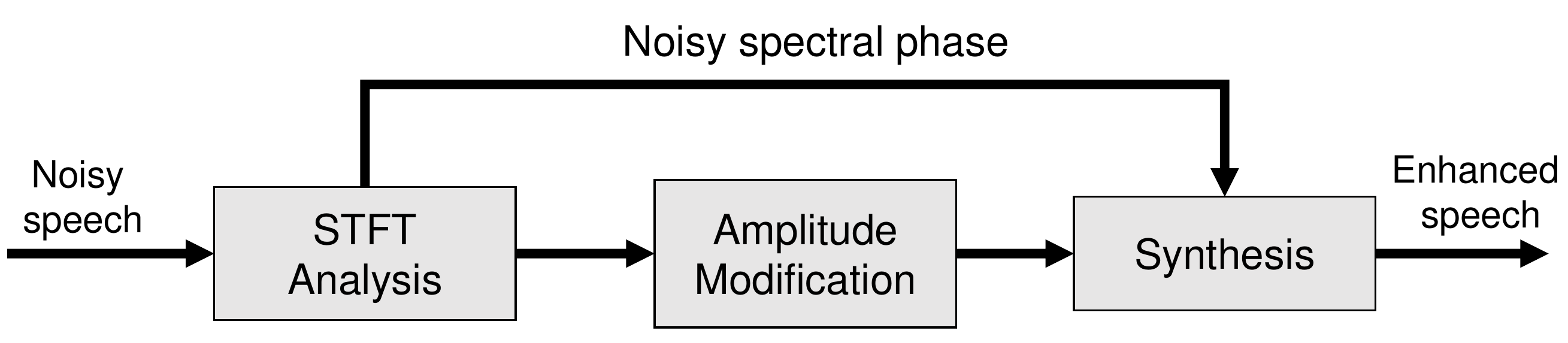}
\caption{Block diagram of conventional speech enhancement algorithm using Analysis Modification Synthesis (AMS) framework.}
\end{figure}
\label{fig2.1}

The present work focuses on the Bayesian estimation approaches of SCSE  based on statistical assumptions about the speech and noise spectral components.  These algorithms mainly follow an Analysis-Modification-Synthesis framework as illustrated in Figure 1. The noisy speech signal is transformed to some spectral domain, more often in short-time Fourier transform (STFT) domain in the analysis stage and then, short-time spectral amplitude (STSA) of the clean speech is estimated by applying a time-varying gain (noise-suppression) function on the observed noisy speech spectrum. Finally, the modified spectrum is synthesized in time domain by using inverse STFT followed by overlap-add synthesis method. Starting from the Ephraim-Malah's original idea of noise-suppression rule based on MMSE estimation of STSA \cite{ephraim1984}, many improvement in this research direction had been suggested by either considering perceptually motivated cost functions such as logarithmic mean square error of spectral amplitude \cite{ephraim1985speech}, weighed Euclidean \cite{loizou2005}, $\beta$-order compression \cite{you2005beta} or assuming better statistical model of clean speech STSA such as generalized Gamma distributions (more-heavy tailed Gaussian distributions) \cite{shin2005statistical}\cite{martin2005speech}\cite{borgstrom2011log} while deriving the Bayesian estimators. Due to earlier belief of phase unimportance in human audio perception \cite{wang1982unimportance}, the baseline Bayesian estimators of STSA have in common that only the speech spectral amplitude is modified, while the noisy phase is left unaltered \cite{ephraim1992statistical}. However, the conclusive results of various recent experimental studies \cite{paliwal2011importance}\cite{gerkmann2015phase}\cite{krawczyk2014stft}\cite{mayer2017impact} have established that the spectral phase information is important for improving both the  speech quality and intelligibility. Hence, it has motivated the subsequent development of different phase estimation algorithms \cite{kulmer2015harmonic}\cite{mowlaee2012phase}\cite{krawczyk2014stft}\cite{mowlaee2012phase}\cite{sugiyama2013phase} and phase-aware statistical STSA estimators \cite{mowlaee2013iterative}\cite{gerkmann2013mmse} for speech enhancement in recent years. Although the phase-only enhancement can improve the perceptual quality (ona n average 0.1-0.5 improvement in Perceptual Evaluation of Speech Quality (PESQ) score) and intelligibility to some extent even at low SNR levels, it is always desirable to find a way to utilize effectively an estimate of STFT phase to further improve the performance of conventional speech enhancement method which modifies only the STFT amplitude in order to take the advantages of both the entities (amplitude and phase) \cite{gerkmann2015phase}. In fact, once an estimate of the clean speech spectral phase is available, the following  approaches can be employed to use the additional phase information for an improved speech enhancement \cite{krawczyk2016mmse}:
\begin{itemize}
\item[(i)] The most straight-forward way is to simply combine the enhanced STSA with the estimated short-time spectral phase (STSP) and reconstruct the modified signal in the time domain. However, if an estimate of the clean speech STFT phase is available, the traditional phase-blind Bayesian estimation approaches, like the MMSE-STSA, are not MMSE optimal any more.
 
\item[(ii)]  Alternatively just like in \cite{gerkmann2013mmse}, a phase-aware Bayesian estimator of the clean speech STSA can be derived, which is optimal in the MMSE sense if the true clean speech spectral phase is given. In practice, however, typically only an estimate of the clean speech spectral phase can be obtained via the phase estimation approaches in [14], [15] or iteratively as proposed in [17].

\item[(iii)] Instead of estimating phase and magnitude separately, they can be jointly estimated in the STFT domain by incorporating the uncertainty in a phase estimate such as complex spectral speech coefficients estimator given an uncertain phase estimate \cite{gerkmann2014bayesian}.

\item[(iv)] Another approach of jointly estimate magnitude and phase is to extend  the phase-sensitive STSA estimator into the loop of an iterative approach that enforces consistency. 
\end{itemize} 
In this work, to take advantage of both the perceptually motivated cost function involving STSAs of estimated and true clean speech and utilizing the prior spectral phase information, we have derived a phase-aware Bayesian STSA estimator. We consider the estimation of the speech STSA using a parametric Bayesian cost
function and speech prior generalized Gamma distribution (GGD) following the approach in \cite{krawczyk2016mmse}. The parameters of the cost function are chosen based on the the characteristics of the human auditory system, namely, the dynamic compressive non-linearity of the cochlea, the perceived loudness theory and the simultaneous masking properties of ear. This type of parameter selection scheme results in more noise reduction while limiting the speech distortion. The derived STSA estimator is optimal in the MMSE sense if the prior phase information is available. In practice, however, typically only an estimate of the clean speech phase can be obtained via employing  different types of spectral phase estimation techniques which have been developed throughout the last few years. In a blind setup, we have evaluated the proposed Bayesian STSA estimator with using different types of standard phase estimation methods available in the literature. Experimental results have shown that the proposed estimator can achieve substantial improvement in performance than the traditional phase-blind approaches as well as the other existing phase-aware Bayesian estimators such as those based on the $\beta$-order minimum mean-square error (MMSE) of the STSA in terms of various objective measures.

The remainder of the paper is organized as follows. After introducing the basic signal model and notation in Section II,  we derive the proposed phase-aware Bayesian estimator for speech enhancement in Section III. In Section IV, Experimental results are presented along with discussion. Finally, Section V concludes the paper.
\section{Signal Model and Notation}
At the input, SCSE system only observe the noisy speech signal, $y(n)$,  which consists of a clean speech, $s(n)$, and the additive noise signal, $w(n)$, that is statistically independent of $s(n)$. The sampled $y(n)$ is split into overlapping segments and each segment is transformed to the Fourier domain after an analysis  window has been applied.  We can assume that the complex STFT coefficients of the noisy speech $Y(p,k)$ are given by an additive superposition of uncorrelated zero-mean speech coefficients  $S(p,k)$ and noise coefficients $W(p,k)$ as
\begin{equation} 
Y(p,k) = S(p,k) + W(p,k)
\end{equation}
where $k$ denotes the acoustic frequency index and $p$ denotes the acoustic frame index. 
In the polar form, the following can be written from (1):
\begin{equation}
R{e}^{j{\theta}_{y}} = A{e}^{j{\theta}_{s}} + {V}{e}^{j{\theta}_{w}}
\end{equation}
where ($R$, $A$, $V$)  represent the short-time spectral amplitudes and (${\theta}_{y}$, $ {\theta}_{s}$, ${\theta}_{w}$) denote the short-time spectral phase components of the noisy speech, clean speech and noise, respectively at the frequency bin $k$ of analysis frame $p$. The frame index and frequency index shall be discarded for better readability.  
We further assume that an initial estimate of clean phase $\tilde{\theta}_s$, i.e., this can be obtained from any phase estimation algorithms. The phase estimation algorithms shall be more elaborately discussed in Section IV. 
\section{Bayesian STSA Estimation given A Clean Spectral Phase Estimate}
The Bayesian STSA estimation problem can be formulated
as the minimization of the expectation of a cost
function $C(A,\hat{A})$ which represents a measure of distance between the true and estimated speech STSAs, denoted respectively
by $A$ and $\hat{A}$. 
The optimal speech STSA estimate in a Bayesian sense can be expressed as 
\begin{equation}
\hat{{A}}^{(o)} = \underset{\hat{A}} {\mathrm{arg min}} \hspace{2 mm} E[C(A,\hat{A})|Y,\tilde{\theta}_s]
\end{equation}
In this work, we have considered a parametric Bayesian estimator whose cost function is given by
\begin{equation}
C(A,\hat{A}) = \bigg{(}\dfrac{A^{\beta}-\hat{A}^{\beta}}{A^{\alpha}}\bigg{)}^{2}
\end{equation}
where $\alpha$ and $\beta$ are two adaptive parameters.  Substituting (4) into (3) and minimizing the expectation, the following can be obtained
\begin{equation}
\hat{{A}}^{(o)}=\Bigg{(}\dfrac{E\{A^{\beta-2\alpha}|Y,\tilde{\theta}_s\}}{E\{A^{-2\alpha}|Y,\tilde{\theta}_s\}}\Bigg{)}^{1/\beta}
\end{equation}
The conditional moments of the form $E\{A^{m}|Y\}$ appearing
in (5) can be obtained as
\begin{equation}
\begin{split}
E\{A^{m}|Y,\tilde{\theta}_s\} \int_{0}^{\infty} \int_{0}^{2\pi} A^{m}p(A,\theta_s|R,\theta_y,\tilde{\theta}_s)d{\theta_s}dA 
\end{split}
\end{equation}
Using Bayes' theorem,
\begin{equation}
\begin{split}
E\{A^{m}|Y,\tilde{\theta}_s\} = \dfrac{\int_{0}^{\infty} \int_{0}^{2\pi} A^{m}p(Y|A,\theta_s)p(A)p(\theta_s|\tilde{\theta}_s)d{\theta_s}dA}{\int_{0}^{\infty} \int_{0}^{2\pi}p(Y|A,\theta_s)p(A)p(\theta_s|\tilde{\theta}_s)d{\theta_s}dA}
\end{split}
\end{equation}
Traditional approaches \cite{ephraim1985speech} model speech spectral (DFT) coefficients as complex Gaussian distribution, corresponding to Rayleigh-distributed spectral amplitudes.
 In this work, we explore the use of generalized Gamma distributed (GGD) speech STSA priors, which are experimentally shown \cite{shin2005statistical} to more accurately approximate empirical histograms of speech ( particularly when the frame-size is less than 100 ms \cite{martin2005speech}),
\begin{equation}
p(A) = \frac{\kappa \lambda^\mu}{\Gamma(\mu)}A^{\kappa\mu-1}\exp(-{\lambda}A^{\mu})
\end{equation}
where $\kappa$ and $\mu$  are known as the shape parameters and $\lambda$ as the scaling parameter \cite{borgstrom2011unified}. $\Gamma(.)$ denotes Gamma function. In order to get a close-form solution, we have set $\kappa$ = 2 which makes (8) to a generalized form of $\chi$-distribution. Based on the second moment of the derived $\chi$-distribution, it can be deduced that the two parameters $\lambda$ and $\mu$ must satisfy the relation $\mu$/$\lambda$ = ${\sigma_s^2}$ \cite{borgstrom2011unified}. So, the distribution of clean speech STSA becomes,
\begin{equation}
p(A) = \frac{2}{\Gamma(\mu)}\Big{(}\frac{\mu}{{\sigma_s^2}}\Big{)}^\mu A^{2\mu-1}\exp(-\frac{\mu}{{\sigma_s^2}}A^{\mu})
\end{equation}
Now assuming the uniform PDF for the speech spectral phase and complex zero-mean Gaussian PDF for the noise spectral coefficients,
\begin{equation}
p(R,\theta_y|A,\theta_s) = \dfrac{R}{\pi\sigma_w^2}\exp\bigg{(}\dfrac{2RA\cos(\theta_s-\theta_y)-R^2-A^2}{\sigma_w^2}\bigg{)}
\end{equation}
As proposed in \cite{gerkmann2014bayesian}, we will model $p(\theta_s|\tilde{\theta}_s)$ as Von-Mises distribution with mean direction $\tilde{\theta}_s$.
\begin{equation}
p(\theta_s|\tilde{\theta}_s) = \dfrac{\exp(\tau\cos(\theta_s-\tilde{\theta}_s))}{2\pi I_o(\tau)} 
\end{equation}
where $\tau$ denotes the concentration parameter and $I_n(.)$ is the $n^{th}$ order modified Bessel function. For an increasing concentration parameter $\tau$, the circular variance of (3.11) decreases, while the mean direction $\tilde{\theta}_s$ corresponds to the mode of the circularly symmetric von-Mises distribution. The von-Mises distribution hence allows us to effectively model the certainty of the available initial phase estimate $\tilde{\theta}_s$ by adequately choosing $\tau$. For large values of $\tau$, $p(\theta_s|\tilde{\theta}_s)$  is strongly concentrated around $\tilde{\theta}_s$. Accordingly, the true clean speech phase $\theta_s$ is likely to be reasonably close to the initial phase estimate $\tilde{\theta}_s$. For small values of $\tau$ on the other hand, $p(\theta_s|\tilde{\theta}_s)$ approaches a uniform distribution, i.e., the initial phase estimate $\tilde{\theta}_s$  yields only little information about the true clean speech phase.
In the following, we now derive various phase-aware parameterized  estimators of the speech amplitude based on different conditions of spectral phase information. 
\subsection{Case-I: Estimation of STSA Given Uncertain Phase}
To find closed-form solution of (3.7), we insert (3.10) and (3.11) into (3.7) and then solve the integral over the speech amplitude using \cite{gradshteyn2000table}, leading to
\begin{equation}
\begin{split}
E\{A^{m}|Y,\tilde{\theta}_s\} & = \Bigg{(}\sqrt{\frac{1}{2}\frac{\zeta}{ \mu + \zeta}\sigma_w^2}\Bigg{)}^m \dfrac{\Gamma(2\mu + m)}{\Gamma(2\mu)}\times \\ & \dfrac{\int_{0}^{2\pi}e^{\nu^2/4}D_{(-2\mu-m)}(\nu)p(\theta_s|\tilde{\theta}_s)d{\theta_s}}{\int_{0}^{2\pi}e^{\nu^2/4}D_{(-2\mu)}(\nu)p(\theta_s|\tilde{\theta}_s)d{\theta_s}}
\end{split}
\end{equation}
where 
\begin{equation}
\nu = -\dfrac{R}{\sigma_w}\Bigg{(}\sqrt{2\frac{\zeta}{\mu + \zeta}}\Bigg{)}\cos(\Delta \theta)
\end{equation}
Here, $\Delta \theta$  denotes the difference between the observed phase $\theta_y$ and the true clean speech phase $\theta_s$, i.e, $\Delta \theta$ = $\theta_y-\theta_s$. The parameters $\gamma$ = $\dfrac{R^2}{E\{V^2\}}$= $\dfrac{R^2}{{\sigma_w^2}}$, and $\zeta$ = $\dfrac{E\{A^2\}}{E\{V^2\}}$= $\dfrac{{\sigma_s^2}}{{\sigma_w^2}}$  are called the $a$ $priori$ and $a$ $posteriori$ SNRs, respectively. $D_{.}(.)$ denotes the parabolic cylinder function. ${\sigma_s^2}$ and ${\sigma_w^2}$ represent power spectral density (PSD) the clean speech and noise, respectively. Unfortunately, there is no known closed-form solution to the phase integral for a von-Mises phase prior. However, since the integral over the phase is limited to -$\pi$ and $\pi$, it can be solved numerically with high precision \cite{gerkmann2014bayesian}.
\subsection{Case-II: Perfectly Known Speech Phase ($\tau$ $\rightarrow$ $\infty$)}
Now assuming that the initial phase estimate yields exactly true clean phase, i.e., $\tau$ = $\infty$, it yields Equation (3.7) as 
\begin{equation}
\begin{split}
{\hat{A}}_{PA} & = \Bigg{(}\sqrt{\frac{1}{2}\frac{\zeta}{\mu + \zeta}\sigma_w^2}\Bigg{)}\times \\   & \Bigg{(}\dfrac{\Gamma(2\mu + \beta - 2\alpha)D_{(-2\mu-\beta+2\alpha)}(\nu)}{\Gamma(2\mu-2\alpha)D_{(-2\mu+2\alpha)}(\nu)}\Bigg{)}^{1/\beta}
\end{split}
\end{equation}
Please note that ${\hat{A}}_{PA}$ in (3.14) does not incorporate any uncertainty in the prior phase information. In practice, however, the initial phase yields only an estimate of the clean speech phase. By choosing, the uncertainty of this estimate is neglected, potentially leading to suboptimal enhancement results for an unreliable initial phase. As the estimator in (3.14) is defined as an estimator of spectral amplitudes only, for signal reconstruction we again use the noisy phase. So, this method does not encounter any artifacts generating from phase modifications. We denote the spectral amplitude estimator in (3.14) as PA-Aud-model STSA estimator.
 \begin{figure*}[t]

 \centering

  \begin{tabular}{cc}


    \includegraphics[scale = 0.38]{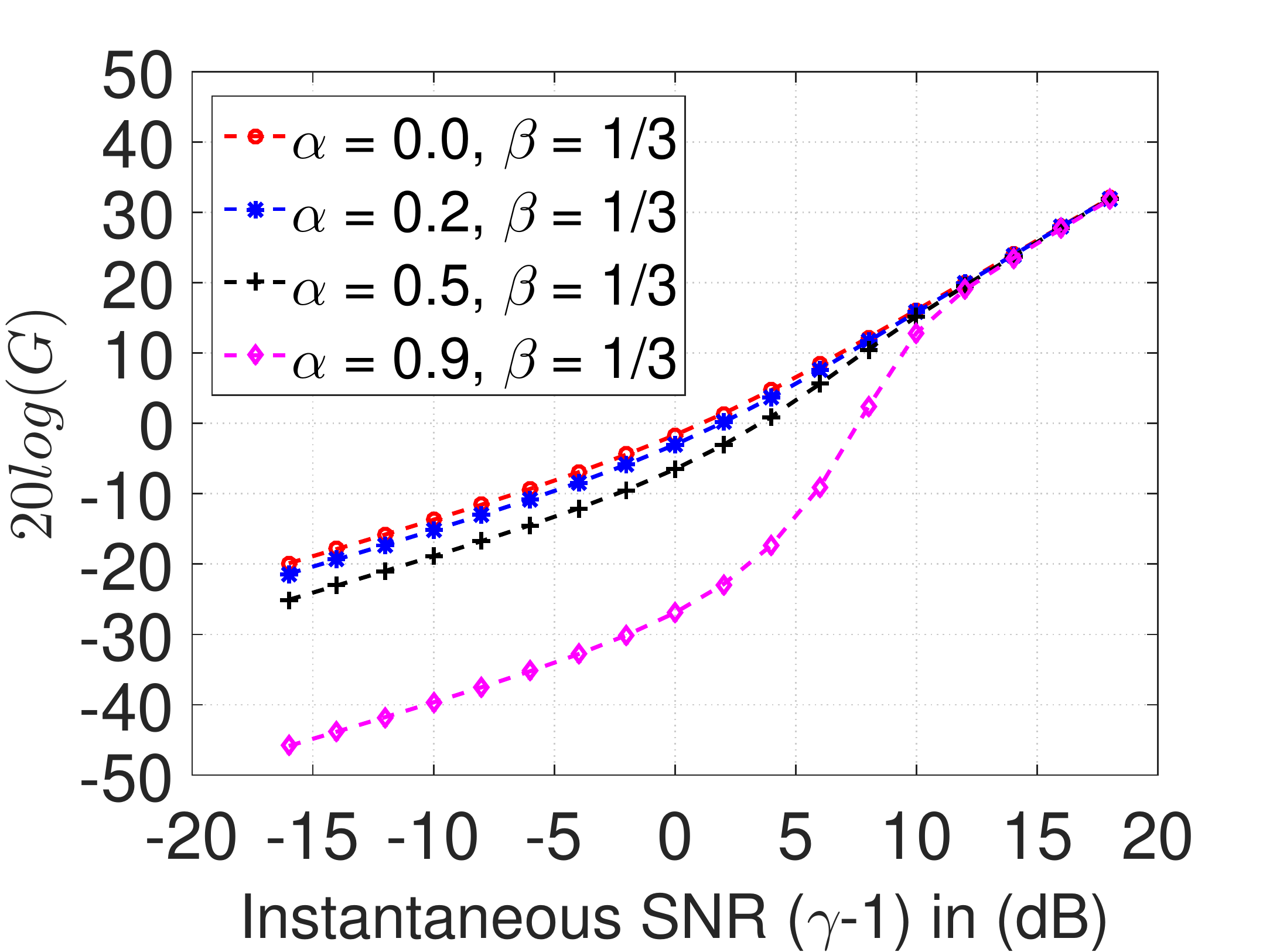} &
    
    \includegraphics[scale = 0.38]{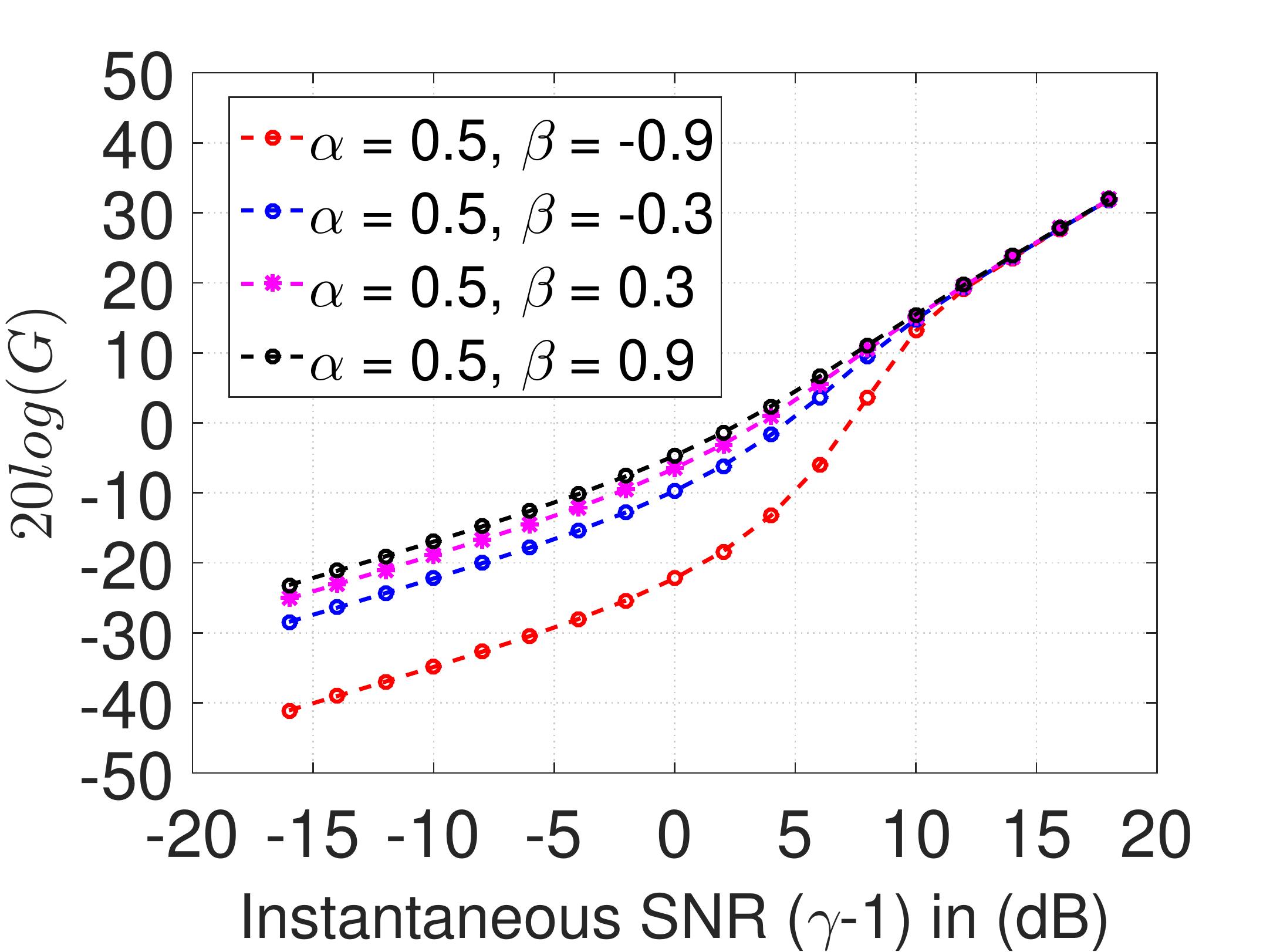} \\

  \end{tabular}
  \label{fig3}\caption{(a) Gain of the parametric STSA estimator: $20\log(G)$ versus instantaneous SNR $(\gamma-1)$ for several values of $\beta$. (b) $20\log(G)$ versus $(\gamma-1)$ for several values of $\alpha$.}
\end{figure*}

\subsection{Case-III: Phase-Blind ($\tau$ = 0)}
For $\tau$ = 0, the von-Mises distribution in (3.11) reduces to a uniform distribution, i.e.,  $p(\theta_s|\tilde{\theta}_s)$ = $\frac{1}{2\pi}$ between $-\pi$ and $\pi$, and zero elsewhere. The estimator (3.7) becomes phase-blind. Hence, the following estimate of STSA can be obtained:
\begin{equation}
\begin{split}
{\hat{A}}_{PB}  & = \Bigg{(}\sqrt{\frac{\zeta}{\mu + \zeta}\sigma_w^2}\Bigg{)}\times \\ & \Bigg{(}\dfrac{\Gamma(\dfrac{-2\alpha + \beta + 2\mu}{2})M(\dfrac{2+2\alpha -\beta -2\mu}{2},1;-\nu')}{\Gamma(-\alpha+\mu)M(1+\alpha-\mu,,1;-\nu')}\Bigg{)}^{1/\beta}
\end{split}
\end{equation}
where
\begin{equation}
\nu' = \frac{\zeta}{\mu + \zeta}\gamma
\end{equation}
In the experiment section, we will denote the spectral amplitude estimator in (3.15) as PB-Aud-model STSA estimator. Here, the index $PB$ stands for $Phase$-$Blind$ as the STSA estimator in (3.15) is indeed insensitive to spectral phase changes. The PB-Aud-model STSA estimator gain depends on the parameters of the cost function (i.e., $\alpha$ and $\beta$ ) as well as on $\zeta$ and $\gamma$. Figure 3.2 presents gain curves as a function of the instantaneous
SNR ($\gamma$-1) for $\zeta$ = 0 dB and several $\alpha$ and $\beta$ values. As can be observed, the estimator’s gain decreases when $\alpha$ increases and increases when $\beta$ increases. It is worth noting that a decrease in the gain will result in more noise reduction but will invariably introduce more speech distortion.

To understand the benefits of the proposed phase-aware estimator, the
input-output curve of (3.14) is given in Figure 3.2 for an a $priori$ SNR of $\zeta$ = 0.5 . To draw conclusions independent of
an absolute signal-scaling, we normalize the input $R$ and the output $\hat{A}$ by $\sigma_w$. In Figure 2, we set the shape parameter $\mu$ = 1, $\alpha$ = 0.5, and $\beta$ = 0.3. The phase information employed by the proposed estimator helps to distinguish if large amplitudes $R/\sigma_w$ $>>$ 1 originate from speech or represent outliers in the noise. For this distinction, conventional estimators only have the a $priori$ SNR $\zeta$ and $R/\sigma_w$ available. Taking the phase into account, we now have additional information for an improved separation of noise outliers
from speech:  if $R/\sigma_w$ is large due to a contribution from speech, then the phase of the noisy speech will be close to the clean
speech phase \cite{vary1985noise}, i.e., $|\Delta \theta|$ $\rightarrow$ 0. Consequently, if $|\Delta \theta|$ = 0, the proposed estimator (top solid line) applies less attenuation and thus less speech distortions than the phase-insensitive STSA estimator. However, if $R/\sigma_w$ is large because of noise outliers, the phase difference $|\Delta \theta|$ is likely to be larger than zero. Employing this larger in (13) and (12) results in an efficient attenuation of noise outliers that is not possible without taking the phase into account. This larger attenuation can be seen in the second, third, and fourth solid line that represent (top to bottom). From these considerations, we see that incorporating the phase provides a novel mechanism to distinguish noise from speech. The
proposed estimator can efficiently reduce undesired noise outliers
while preserving the speech signal.
\begin{figure}[]
  \centering
  \includegraphics[scale = 0.32]{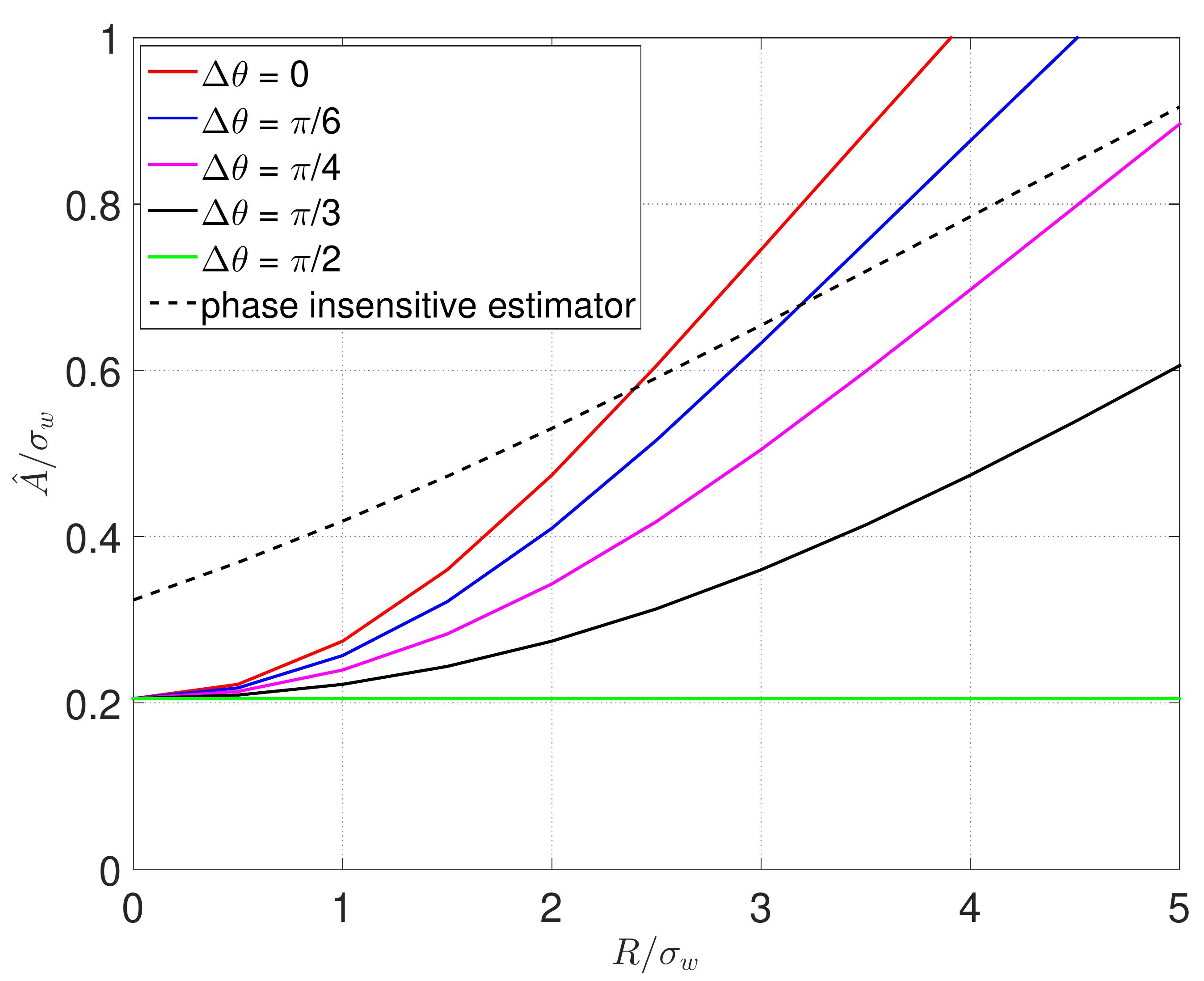}
  \caption{Input-output curve of the proposed estimator (7) for ,
and . The solid lines are the results for different values for
, namely (top to bottom)}
  \label{fig1}
\end{figure}

\subsection{Selection of Parameters: $\alpha$ and $\beta$}
\begin{figure}[]
  \centering
  \includegraphics[scale = 0.4]{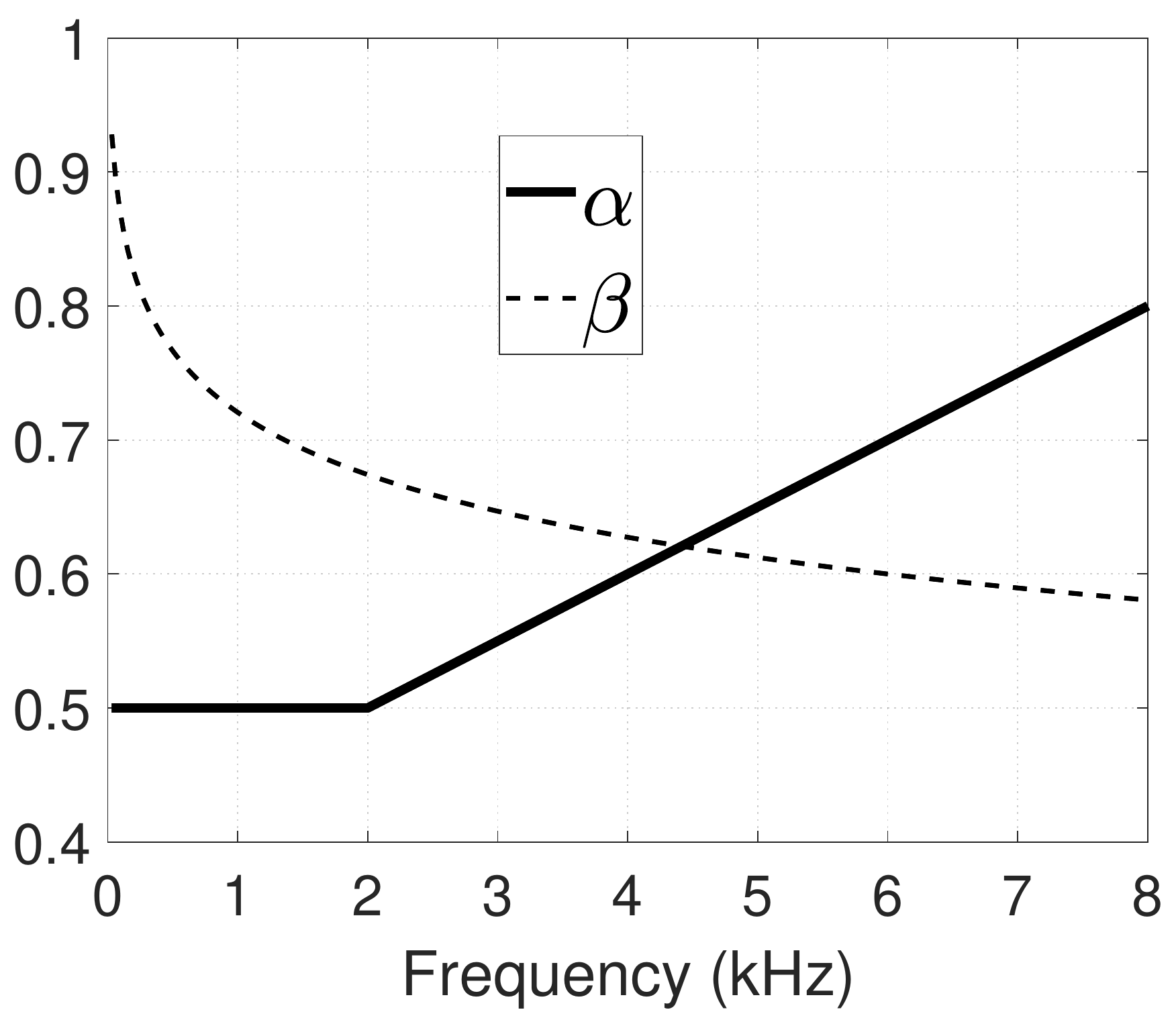}
  \caption{$\alpha$ and $\beta$ variations with frequency}
  \label{fig1}
\end{figure}
The parameter values of speech enhancement algorithms
have been chosen in the past based on frame SNR such that
a higher gain is obtained for higher SNR and vice versa [10],
[16]. This had the effect of removing less noise at higher SNR
to prevent speech distortion and more noise at low SNR.

In spite of using fixed values, we have exploited the characteristics of human auditory system to select the appropriate  values of  $\alpha$ and $\beta$. The value of $\alpha$ can be chosen by taking advantages of the masking properties of human ear. It is always desirable for a STSA estimator to favour a more accurate estimation of smaller STSA since they are less likely to mask noise remaining in the clean speech estimate. Since most of the speech energy is located at lower frequencies, higher frequencies should contain mainly small STSA \cite{formby1982long}. Therefore, it would be relevant to further increase the weights of the smaller STSA in the cost function for higher frequencies. This can be done by increasing $\alpha$  for higher frequencies as follows
\begin{equation}
{\alpha}_{k} = \begin{cases} {\alpha}_{low}, & {f}_{k} \leq 2kHz \\ \dfrac{({f}_{k}-2000)({\alpha}_{high}-{\alpha}_{low})}{\frac{{f}_{s}}{2}-2000}+{\alpha}_{low}, & otherwise \end{cases}
\end{equation}
where parameters, ${\alpha}_{high}$ = 0.8 and ${\alpha}_{low}$ = 0.2 are set empirically. $f_k$ denotes the frequency in Hz corresponding to spectral component $k$, i.e., $f_k$ = $kf_s/N$, $f_s$ denotes sampling frequency. On the other hand, by considering the non-linear compression in the perception of loudness in the human cochlea, we can select the $\beta$ value as follows,
\begin{equation}
{\beta}_{k} = \Bigg{[}\dfrac{\log_{10}(\frac{f_k}{Q}+l)}{\log_{10}(\frac{f_s}{2Q}+l)}\Bigg{]}({\beta}_{high}-{\beta}_{low}) +  {\beta}_{low}
\end{equation}
where $Q$ = 16.54 is an empirical constant relevant to the $tonotopic$ $mapping$ of basilar membrane. We have set empirically: ${\beta}_{low}$ = 1 and ${\beta}_{high}$ = 0.2.  The variation of $\alpha$ and $\beta$ with the acoustic frequency is shown in Figure 3.4.
\section{Experimental results}
To evaluate the performance of the proposed auditory model based phase-aware Bayesian STSA estimator, we have synthetically generated noisy speech stimuli by adding noise to 100 clean utterances taken from TIMIT database \cite{garofolo1993darpa}, sampled at $f_s$ = 16 kHz. The utterances comprise of 10 male and 10 female speakers. Three types of noise instances: babble, factory and pink noise, are taken from NOISEX-92 database \cite{varga1993assessment}. All these noise samples are non-stationary in nature. In addition, we considered another noise type speech-shaped noise (SSN) which is stationary in nature and produced synthetically by filtering a Gaussian white noise sequence through a 12$^{th}$ order all-pole filter with coefficients found from Linear Predictive Coding (LPC) analysis of 20 randomly selected Harvard sentences (IEEE corpus)\footnote{http://www.cs.columbia.edu/hgs/audio/harvard.html}. Each utterance  is mixed with the aforementioned noise instances at 6 SNR levels ranging from -5 dB to 15 dB (with 5 dB step). To obtain desired SNR, the noise level is adjusted based on `active speech level' according to ITU-T P.56 \cite{ITU-TASL}. For analysis and synthesis, we have used Hanning windows of 32 ms with an overlap of 50\%. To increase the perceptual quality of the enhanced signal, we further limit the maximum attenuation in each time-frequency point to -15 dB. We have used an unbiased MMSE-based noise power estimator \cite{gerkmann2012unbiased} to determine the noise spectral variance ${\sigma_w^2}$. Instead of using voice activity detector (VAD), this estimator utilizes soft signal presence probability (SPP) with fixed priors for updating the noise estimate which makes the estimator unbiased and computationally less complex \cite{gerkmann2012unbiased}. The $a$ $priori$ SNR is estimated by the two stage decision-directed approach \cite{plapous2006} with a smoothing factor of 0.98.
\subsection{Instrumental Evaluation metrics}
Although subjective listening test is still considered to be the ultimate criterion for evaluating the speech enhancement performance, it is very time consuming and difficult to standardize. Therefore, we relied on two objective measures: the perceptual evaluation of speech quality (PESQ) and the short-time objective intelligibility (STOI),  to predict the perceived speech quality and the speech intelligibility respectively. Both of them are highly correlated with the subjective listening test. PESQ is an  International Telecommunications Union (ITU-T P.862) standardized intrusive objective measure which predicts the perceived quality of a test utterance by yielding a score in the range [1, 4.5], where a higher score indicates better quality. PESQ score is highly correlated with MOS (mean opinion score) in the subjective listening test. The original PESQ algorithm \cite{recommendation2001perceptual} was developed for narrow band telephonic speech (8 kHz sampling rate). In order to extend the application of PESQ to systems such as wide-band (16 kHz) telephony and speech codecs, a wide-band extension to the PESQ measure has been proposed in ITU-T P.862.2 \cite{rec2005p} and are widely used in the evaluation of speech enhancement algorithms. On the other hand, STOI is  an intrusive metric, mainly computed based on a correlation coefficient between the temporal envelopes of the time-aligned reference (clean) and processed (enhanced) speech signal in short-time overlapped segments \cite{taal2011algorithm}. STOI measure results in a score from 0 to 1, where a higher score indicates better intelligibility. 

\begin{figure*}[htp!]

 \centering

  \begin{tabular}{cc}


    \includegraphics[scale = 0.2]{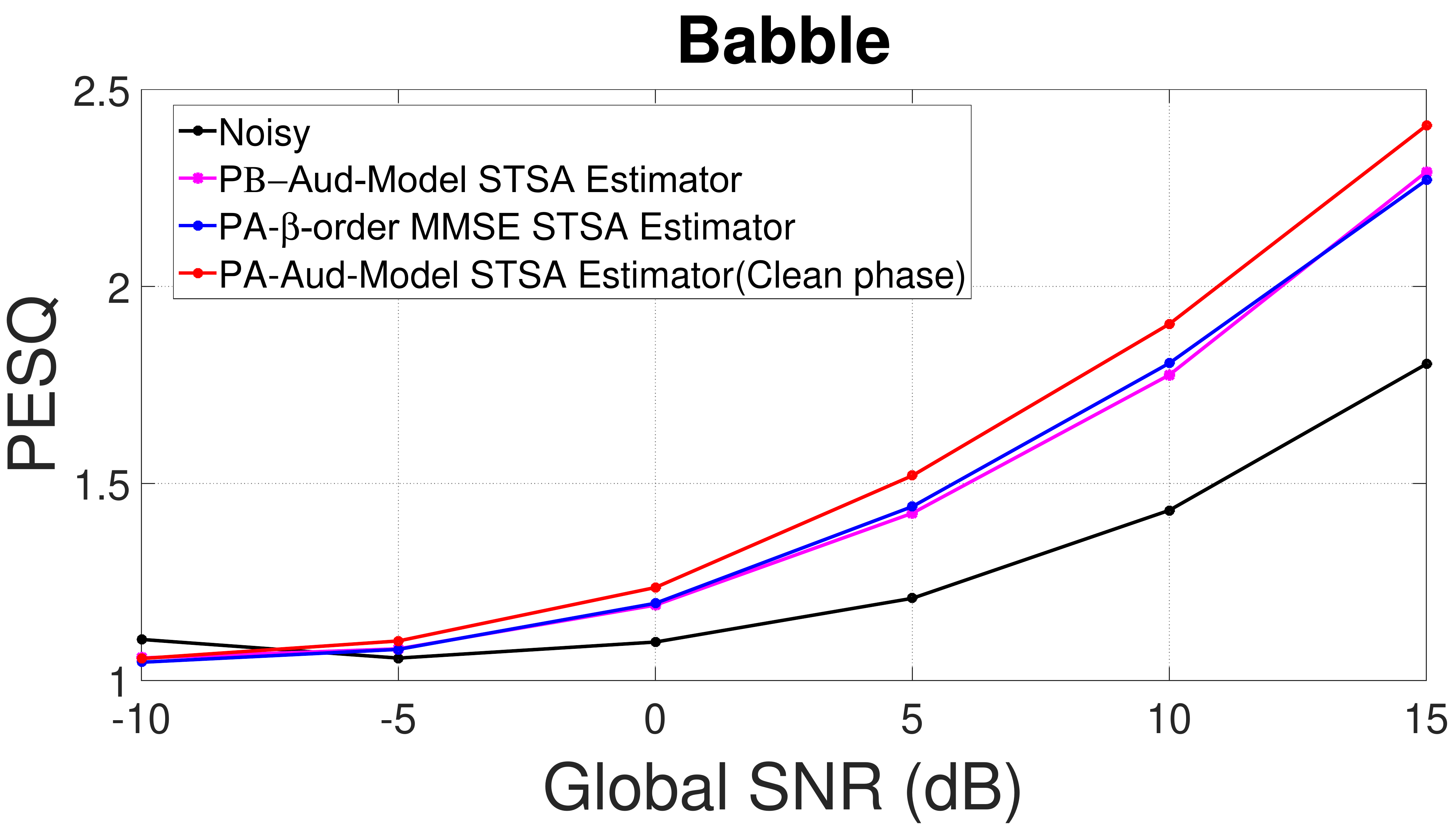} &

    \includegraphics[scale = 0.2]{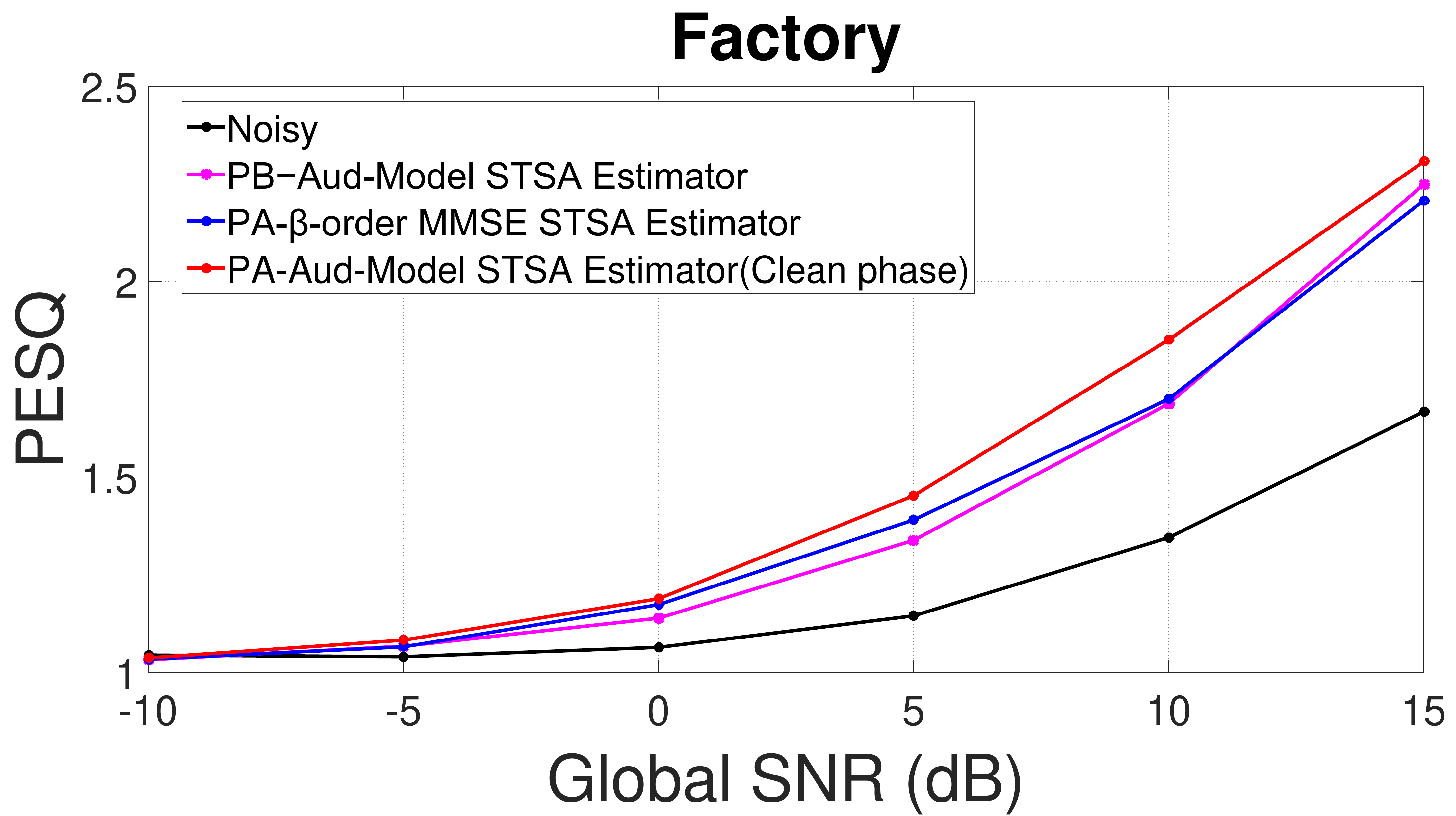} \\

    \includegraphics[scale = 0.2]{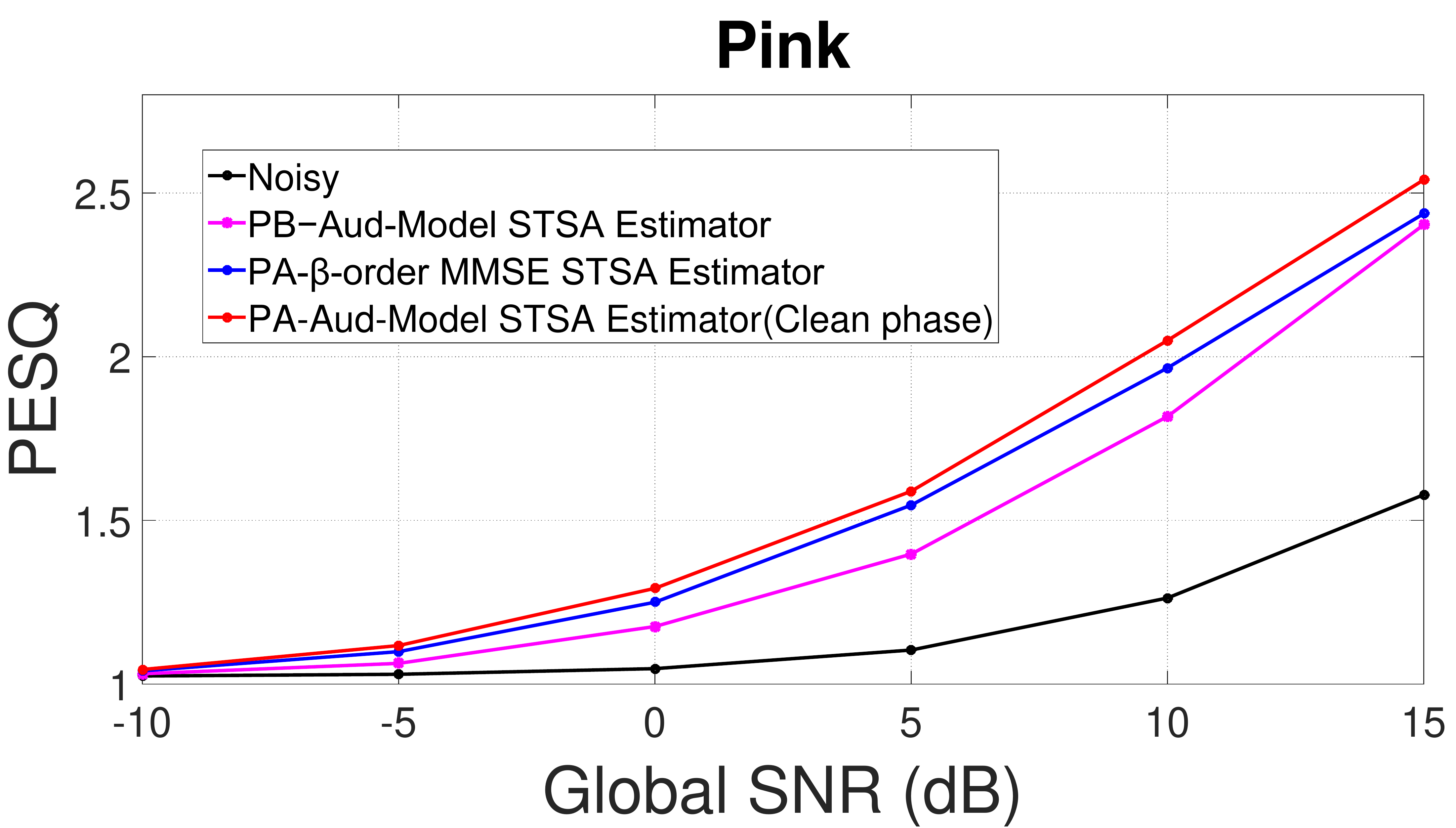} &
    
    \includegraphics[scale = 0.2]{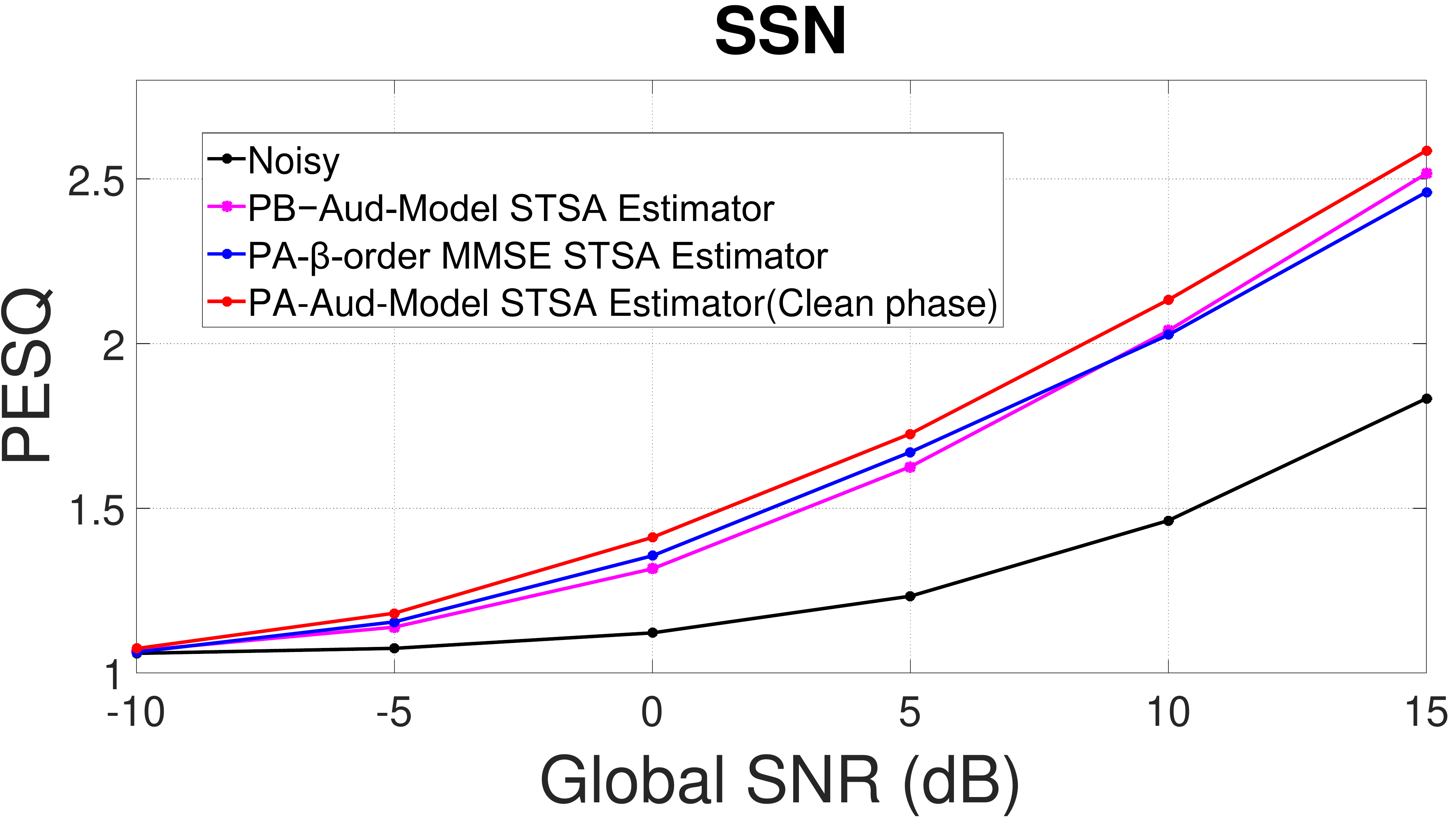} \\

  \end{tabular}
  \label{fig4}\caption{Mean PESQ scores at different input SNR levels in various noise scenarios: (a) Babble noise (b) Factory noise (c) Pink noise (d) SSN noise.}
\end{figure*}

\subsection{Evaluation under Clean Speech Scenario}
We have studied the performance of the proposed auditory model based phase-aware Bayesian estimator (denoted by PA-Aud-model STSA estimator, Eq. (3.14)) and compared it to the phase-blind auditory model based spectral magnitude estimator (as in equation 3.15), denoted by PB-Aud-model STSA estimator, and the phase-aware $\beta$-order MMSE STSA estimator \cite{gerkmann2013mmse} with $\beta$ = $\mu$ = 0.5.  In this experiment, the clean spectral phase $\theta_s$ (oracle information) has been employed to estimate the clean speech STSA. All other parameters such as noise spectral variance ($\sigma_w^2$) and $a$ $priori$ SNR ($\zeta$) are estimated from the noisy microphone signal $Y$. Please note that all phase-blind estimators, as well as the phase-aware amplitude estimators use the noisy phase for signal reconstruction A completely blind approach is presented in the next section, where $\tilde{\theta}_s$ is estimated from the noisy observation $Y$. The average PESQ scores of noisy speech and the enhanced utterance processed by the proposed approach and the other baseline techniques are illustrated in Figure 3.5. It is quite evident from the result that the proposed PA-Aud-model STSA estimator outperforms both the $\beta$-order phase-aware Bayesian estimator and the PB-Aud-model STSA estimator at all SNR levels in all noise conditions. The proposed PA-Aud-model based STSA estimator achieves  on an average 20\% improvement in Babble noise and 30\% improvement in factory noise over the phase-blind STSA estimator (PB-Aud-model) for SNR levels except -10 and -5 dB.  It highlights the importance of considering spectral phase information while deriving the Bayesian STSA estimator. The benefit of the phase-aware
estimators, however, is the largest for non-stationary
noises,  in terms of PESQ, where the additional phase
information allows for a better suppression of noise outliers,
like babble and factory noise. Likewise, the proposed PA-Aud-model Bayesian STSA estimator achieves on an average 15\% improvement over $\beta$-order phase-aware Bayesian estimator for SNR values $\geq$ 0 dB in almost all noise scenarios. It shows that it is indeed advantageous if the  parameters ($\alpha$ and $\beta$) of the STSA estimator are chosen based on the characteristics of the human auditory system. The average STOI improvement ($\Delta$ STOI) for the similar conditions are illustrated in Figure 3.6. We observed that the proposed phase-aware Bayesian estimator shows similar kind of superiority in improving the STOI score while compared to the other baseline methods. The proposed estimator (PA-Aud-model STSA estimator) achieves 30\% improvement compared to $\beta$-order phase-aware MMSE-STSA estimator and 20 \% compared to PA-Aud-model Bayesian STSA estimator.  So, the better performance in terms of STOI implies the benefit of the phase-aware estimators in an increased noise reduction, especially in non-stationary noises, while the speech component is preserved.
\begin{figure*}[]

 \centering

  \begin{tabular}{cc}


    \includegraphics[scale = 0.2]{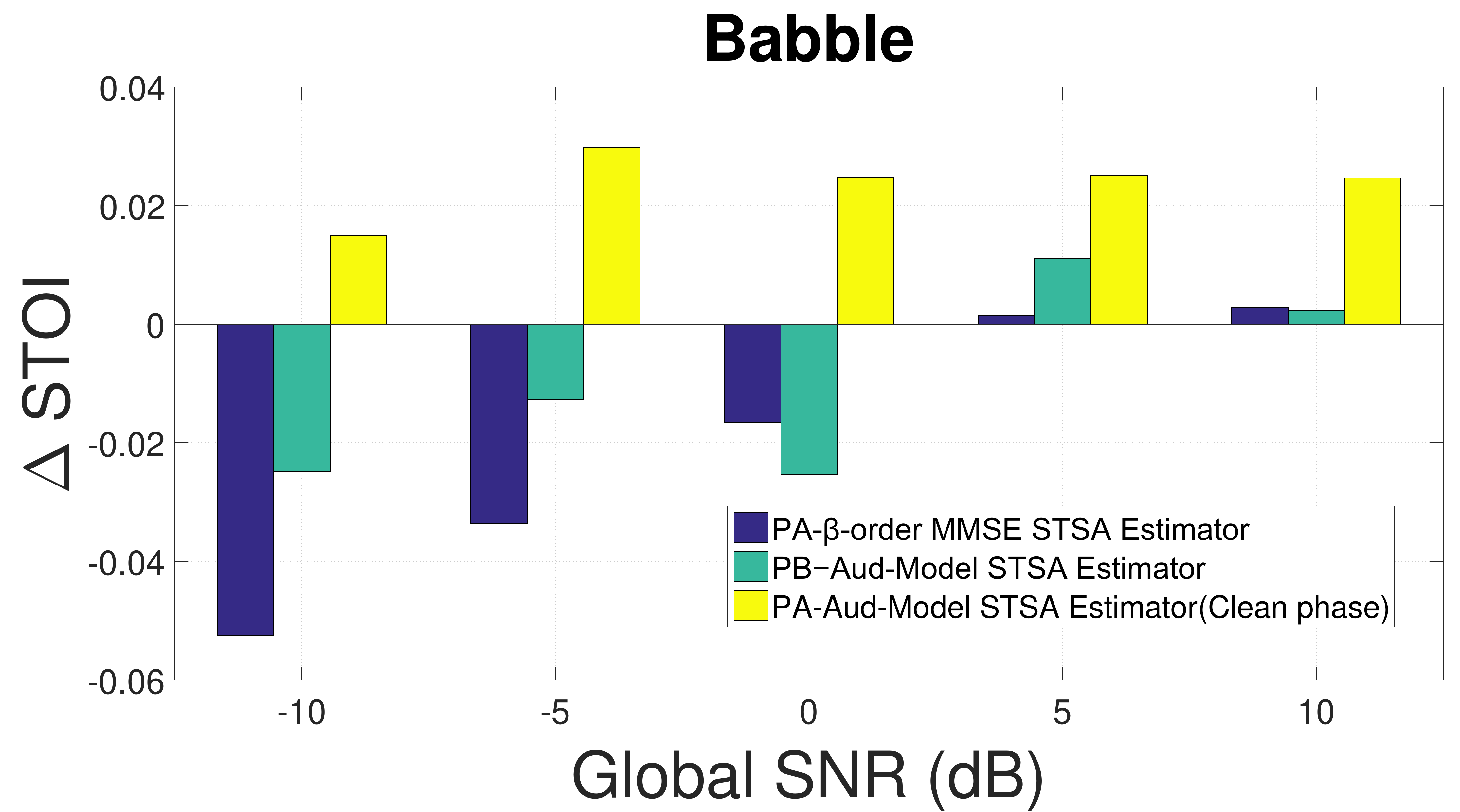} &

    \includegraphics[scale = 0.2]{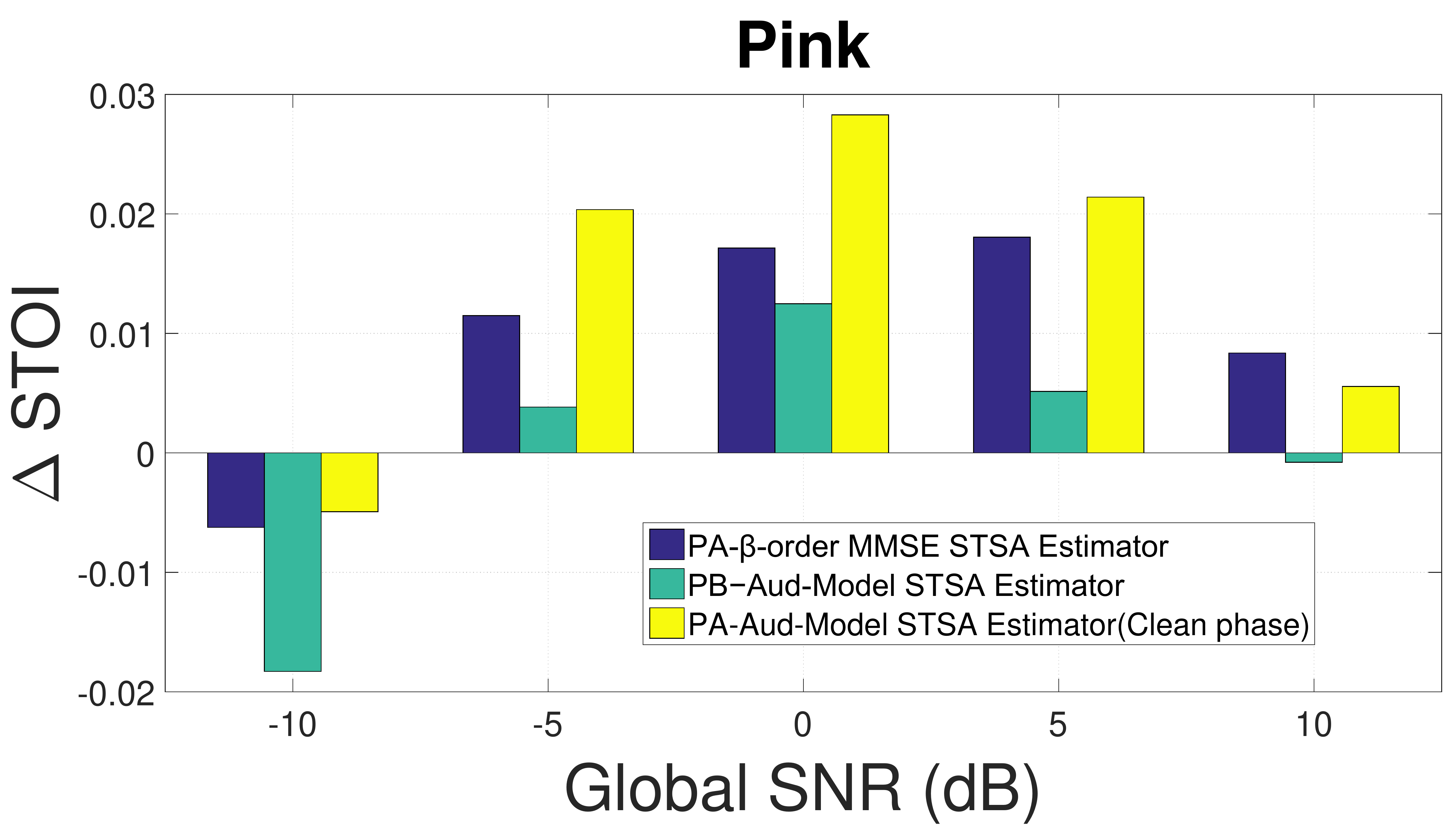} \\

    \includegraphics[scale = 0.2]{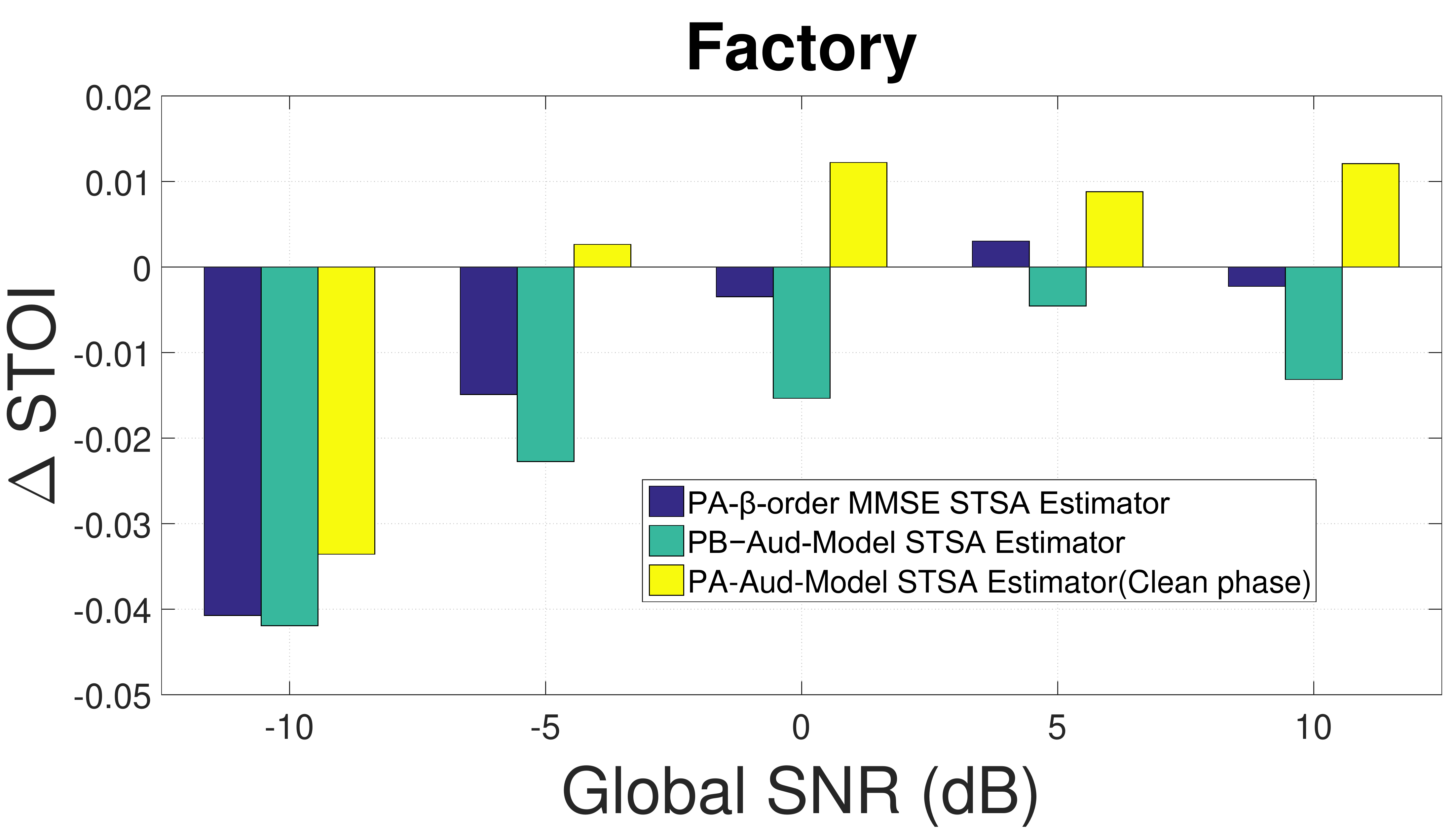} &
    
    \includegraphics[scale = 0.2]{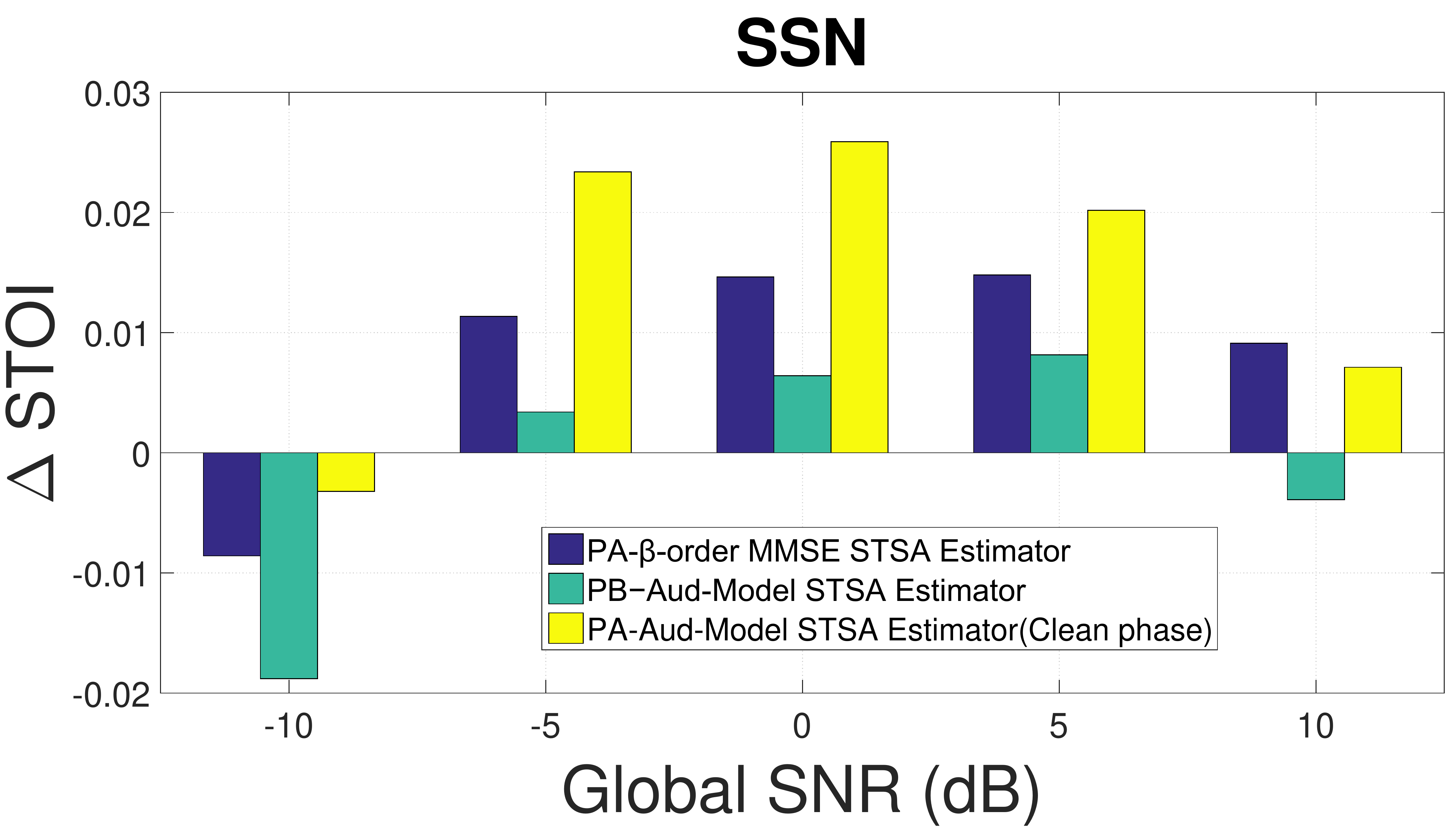} \\
    
  \end{tabular}
  \label{fig4}\caption{Mean STOI score improvement at different input SNR levels for various noise conditions: (a) Babble noise (b) Factory noise (c) Pink noise (d) SSN noise.}
\end{figure*}



\begin{figure*}[]

 \centering

  \begin{tabular}{cc}


    \includegraphics[scale = 0.2]{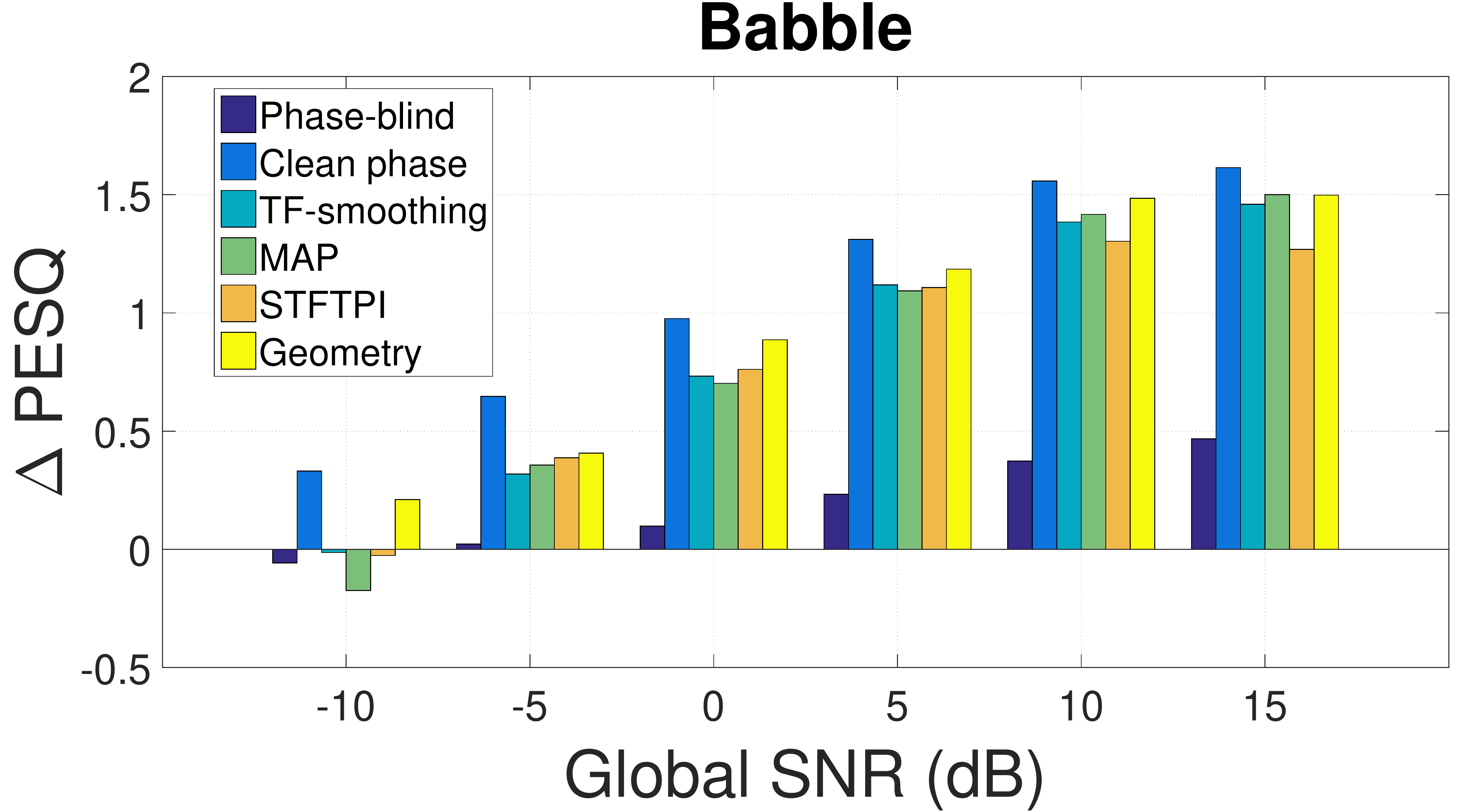} &

    \includegraphics[scale = 0.2]{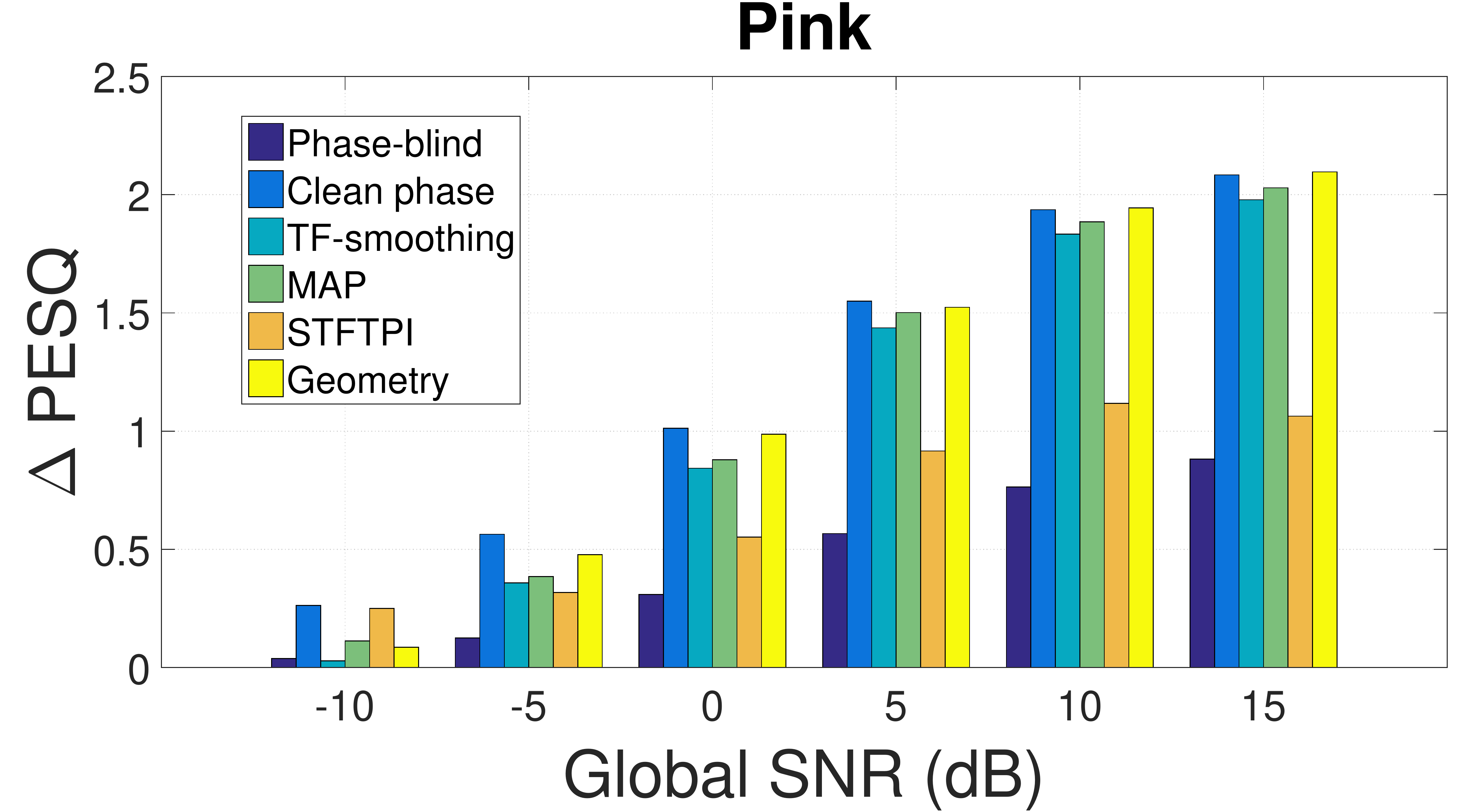} \\

   \includegraphics[scale = 0.2]{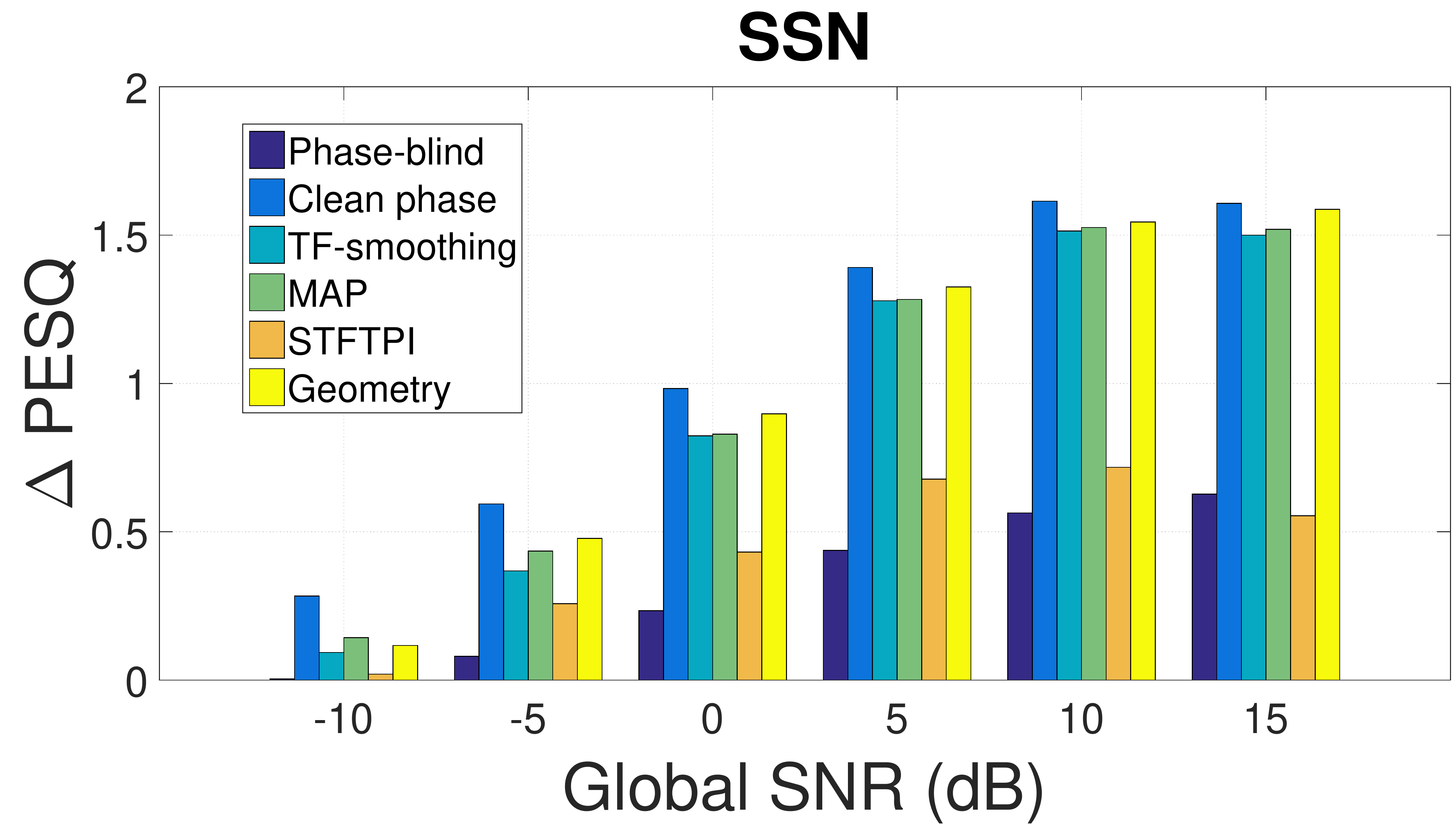} &
   
   \includegraphics[scale = 0.2]{del_pesq_ssn_pa.pdf} \\

  \end{tabular}
  \label{fig4}\caption{Experimental evaluation in completely blind setup. Mean PESQ score improvement at different input SNR levels for various noise conditions: (a) Babble noise (b) Factory noise (c) Pink noise (d) SSN noise. Results are illustrated for various phase estimation methods.}
\end{figure*}


\subsection{Evaluation under Blind Scenario}
In the last section, we have shown the superiority of the proposed phase-aware Bayesian STSA estimator in solving the speech enhancement task over the other existing methods. However, the evaluation has been done based on oracle scenario, i.e, we have assumed that the clean speech spectral phase is available, but in practice, we have to estimate the spectral phase. Therefore, we shall now evaluate the performance of the PA-Aud-model STSA estimator in completely blind scenario, where the spectral phase has to be estimated from the noisy observations. In the last few years, various methods of phase estimation in the context of SCSE have been introduced and these techniques have shown that speech enhancement based on the phase-only modification is more promising in storing the speech intelligibility at different SNR levels, even at a low SNR condition. In the following, we will describe briefly each of these phase estimation algorithms which have been considered for the assessment:
\begin{itemize}
\item \textbf{Harmonic model-based STFT phase reconstruction:} This method utilizes the harmonic model of voiced speech to restore the phase structure of clean voiced speech segments from only the noisy observations as given in \cite{krawczyk2014stft}. Voiced speech with fundamental frequency ${f}_{0}$ can be expressed as the summation of each harmonic sinusoidal component with harmonic index $h$, amplitude $2{A}_{h}$, time domain phase ${\phi}_{h}$, sampling frequency ${f}_{s}$ and harmonic-frequency ${f}_{h}=({f}_{0}+1)h$ :
\begin{equation}
{s}_{v}(n)\approx \sum_{h=0}^{H-1}2{A}_{h}\cos (2\pi n\frac{{f}_{h}}{{f}_{s}}+{\phi}_{h}) 
\end{equation}
where $H$ represents the total number of harmonics. 
It is assumed that each STFT band $k$ is dominated by only one harmonic frequency which is closest to that band.   
For a single frequency component, the clean spectral phase can be estimated using the recursive relation given by
\begin{equation}
\hat{\theta}_{s_{v}}(j,k)=\hat{\theta}_{s_{v}}(j-1,k)+2\pi{{f}_{h}}^k\dfrac{L}{{f}_{s}}
\end{equation} 
where ${{f}_{h}}^k$ is the dominant harmonic frequency in STFT band $k$. For the initialization, noisy phase is used here. After estimating the phase of band $k$ according to (3.20), the spectral phases of adjacent bands of the same frame can be estimated as
\begin{equation}
\hat{\theta}_{s_{v}}(j,k+i)=\hat{\theta}_{s_{v}}(j,k)-{\theta}_{w}(k-{{f}_{h}}^k \frac{N}{{f}_{s}}) + {\theta}_{w}(k+i-{{f}_{h}}^k \frac{N}{{f}_{s}})
\end{equation}
where $ {\theta}_{w} $ is the phase response of spectral analysis window, used for STFT analysis.
Using this harmonic model, the clean spectral phase structure of voiced segments can be estimated, although one needs to know $a$ $priori$ the fundamental frequency ${f}_{0}$ for each segment. The fundamental frequency ${f}_{0}$ is estimated with the noise-robust fundamental frequency PEFAC  \cite{gonzalez2014pefac} estimator. Please note, in the experimental results, we have denoted this method as $STFTPI$ (STFT phase improvement).
\item \textbf{Spectral geometry based method:} This method was first proposed for single-channel source separation problem \cite{mowlaee2012phase}. Later on, it was effectively used for SCSE task \cite{suman_iet} where the phase estimation is possible if $a$ $priori$ knowledge of signal spectra is given. Along with the properties of signal geometry, the proposed method use an additional constraint on group delay deviation in order to solve the ambiguity in the phase estimation problem. Please refer to \cite{mowlaee2012phase} for more details about this method.
\item \textbf{Phase decomposition and time-frequency smoothing:} In \cite{mowlaee2015harmonic}, the authors have proposed a phase estimation method based on phase decomposition and time-frequency smoothing filters. The computational steps of method is summarized in Algorithm 1.
\begin{algorithm}[h!]
\scriptsize
\renewcommand{\arraystretch}{1.2}
 \caption{Phase-estimation}
 
  {\textbf{Inputs:} Input noisy speech signal: ${y}(n)$.

  {\textbf{Output:} Estimated spectral phase of clean speech: $\tilde{{\phi}}_{s}(k,j)$.} 
 \begin{algorithmic}[1]    
     \State \textbf{Analysis stage:} The input signal ${y}(n)$ is segmented by using a pitch-synchronous segmentation \cite{degottex2014uniform} based on harmonic model. This segmented signal ${y}_{w}(n,j)$ can be modelled as the sum of harmonics with amplitude ${A}_{h}$ and harmonic phase $\phi(h,j)$ as:

\begin{equation}
{y}_{w}(n,j)\approx \sum_{h=1}^{H}{A}_{h}\exp (in\omega_{o}(j))\exp (i\phi(h,j))  
\end{equation}
where $H$ is the number of harmonics and $\omega_{o}(j)=2\pi f_{0}(j)/f_{s}$, where $f_{0}(j)$ and $f_{s}$ are the fundamental frequency at frame $j$ and sampling frequency respectively. 
      \State \textbf{Phase-decomposition:} The instantaneous harmonic phase $\phi(h,j)$ is decomposed into two parts as:
\begin{equation}
\phi(h,l)=\phi_{lin}(h,j)  +  \phi_{uw}(h,j)
\end{equation}
where the linear phase $\phi_{lin}(h,j)$  depends on the fame-shift as well as on the fundamental frequency; unwrapped phase components $\phi_{uw}(h,j)$ (minimum-phase + dispersion phase) contains the stochastic information of underlying harmonic phase \cite{agiomyrgiannakis2009wrapped}. Since linear phase part is quite deterministic, so in order to recover the clean speech phase-spectrum, the unwrapped phase is modelled and smoothed to reduce the large variance caused by noise contamination. 
      \State \textbf{Phase-modelling based on circular statistics:} Harmonic phase has different variances depending on SNR. At voiced frames and at high SNR regions, the unwrapped phase is zero-variance (Dirac-delta distribution) and at low SNR regions, the phase distribution will have a large variance. To capture this both scenarios, the unwrapped phase is modelled by circular-statistics based on von Mises distribution as in \cite{mowlaee2015harmonic}.
      \State \textbf{Smoothing:} A spectro-temporal smoothing technique as proposed in \cite{mowlaee2015harmonic} is finally employed in order to reduce the undesirable effects of the additive noise.
      \State \textbf{Phase-combination:} The smoothed unwrapped phase is recombined with the linear-phase part to produce an enhanced instantaneous phase $\tilde{{\phi}}_{s}(h,j)=\phi_{lin}(h,j)+\hat{\phi}_{up}(h,j)$ and subsequently transforming it back into STFT domain $\tilde{{\phi}}_{s}(k,j)$.      
\end{algorithmic}
 }
\end{algorithm}
\item \textbf{Maximum A Posteriori phase estimator:} In \cite{kulmer2015harmonic}, the authors derived a MAP estimator of harmonic phase assuming a von Mises distribution phase prior. Unlike aforementioned methods, this method is less sensitive to fundamental frequency and relies on no voicing state decision which arguably is erroneous at low signal-to-noise ratios and non-stationary noise.
 
\end{itemize}    
%
%
%
%
%
%
%
%
%
%
%
%
%
%
%
%
%
%
%
%
%
%
%


\begin{figure*}[]

 \centering

  \begin{tabular}{cc}


    \includegraphics[scale = 0.2]{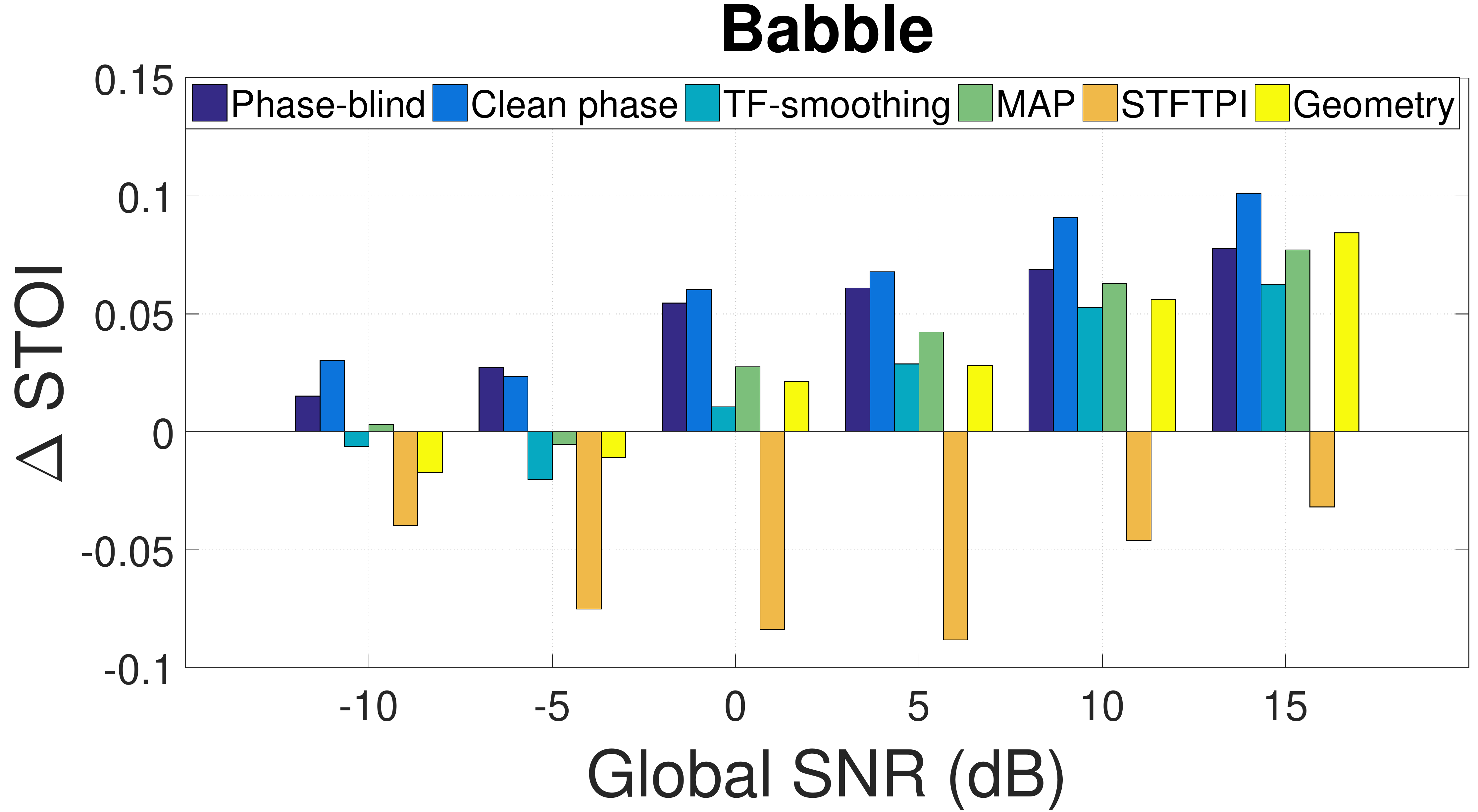} &

    \includegraphics[scale = 0.2]{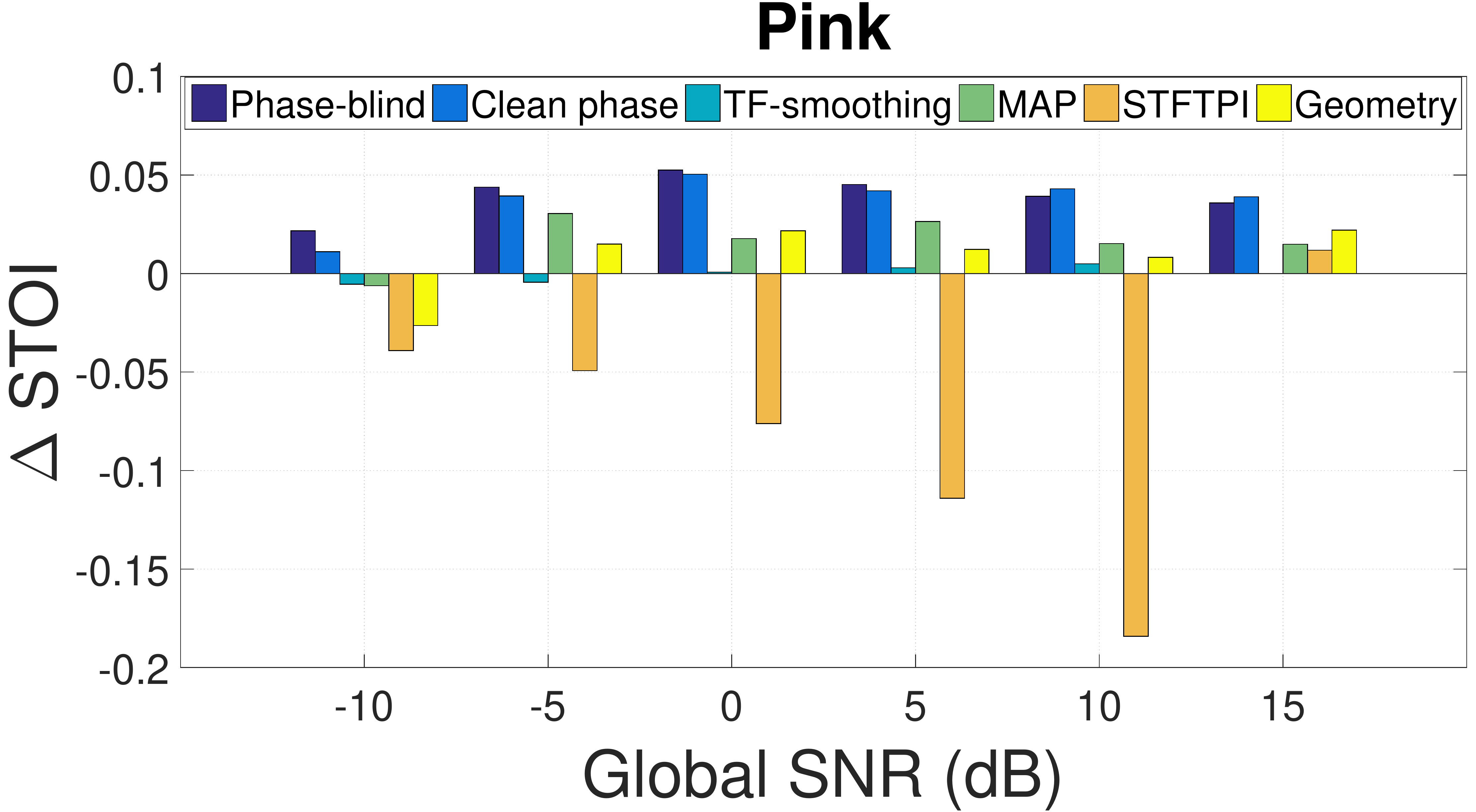} \\
    
    \includegraphics[scale = 0.2]{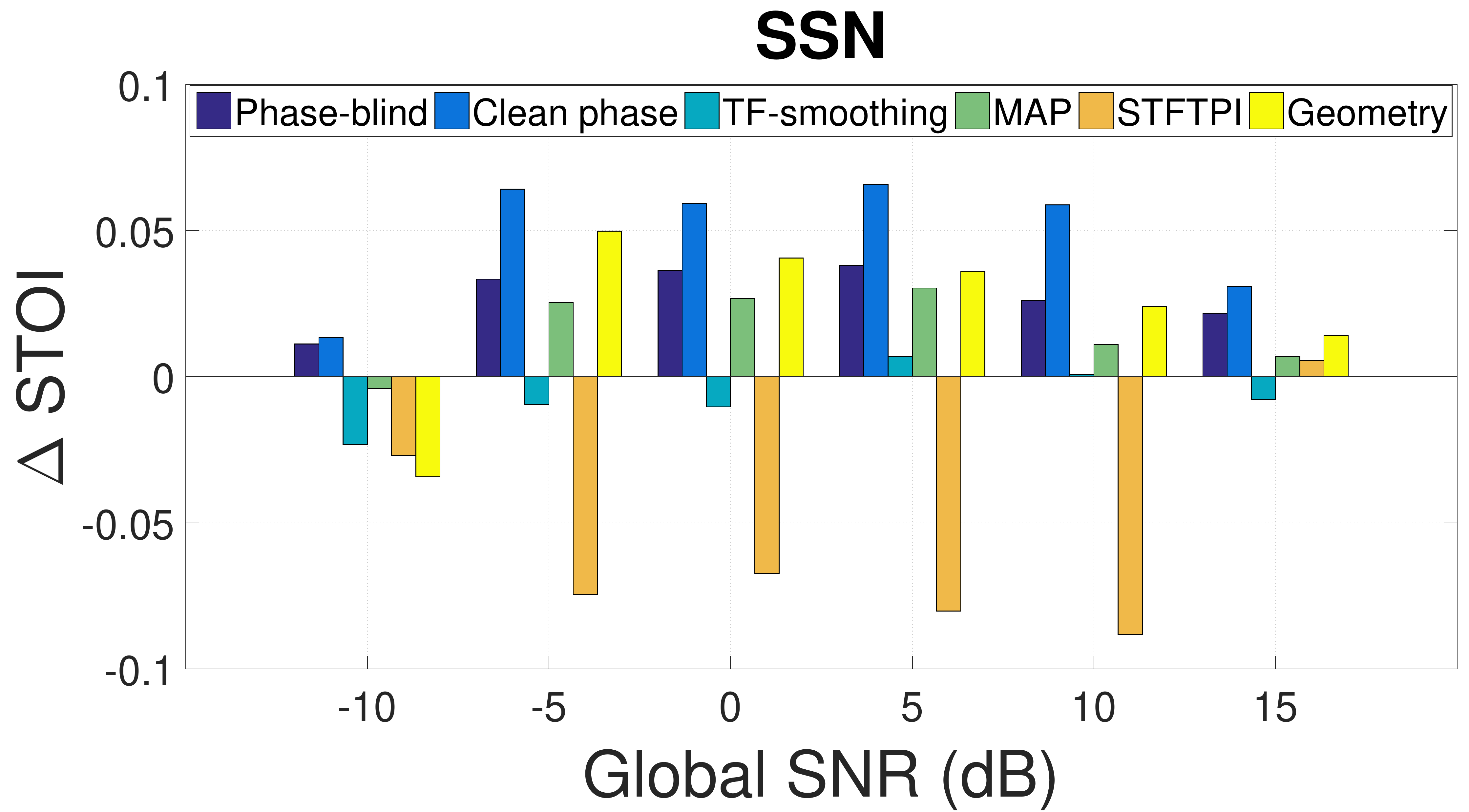} &
    
    \includegraphics[scale = 0.2]{del_stoi_ssn_pa.pdf} \\

  \end{tabular}
  \label{fig4}\caption{Experimental evaluation in completely blind setup. Mean STOI score improvement at different input SNR levels for various noise conditions: (a) Babble noise (b) Factory noise (c) Pink noise (d) SSN noise. Results are illustrated for various phase estimation methods.}
\end{figure*}


Next, each of these phase estimation methods are utilized in the proposed PA-Aud-model based Bayesian STSA estimator. Each of the utterances are processed under total 36 conditions (6 different methods $\times$ 6 SNR levels). Figure 3.7 and Figure 3.8 illustrate a comparison of the average PESQ improvement ($\Delta$ PESQ) and STOI improvement ($\Delta$ STOI) respectively for 6 SNR levels. $phase$-$blind$ method in the legend of the bar graphs indicates the PB-Aud-model Bayesian STSA estimator (please refer to Eq. (3.15)) when spectral phase distribution is considered to be uniform. Whereas, $clean$ $phase$ indicates  the PA-Aud-model Bayesian STSA estimator with the clean spectral phase (oracle case). So, in fact, the 'clean phase' scenario indicates the possible upper limit of performance improvement when spectral phase information is exploited in the proposed auditory model based Bayesian STSA estimator. It is worth noticing that the proposed PA-Aud-model Bayesian STSA estimator which performs the best with the geometry based phase estimation method, while it with STFTPI method shows the worst performance. In case of babble and pink noise, the performance of the geometry based and MAP phase estimator are quite similar with the proposed PA-Aud-model in terms of PESQ and STOI. The PA-Aud-model  with STFTPI method shows comparatively poor performance. Even undesirable artifacts (buzzyness) have been produced in the reconstructed speech.  The possible reason of poor performance of STFTPI is that its applicability is restricted in voiced regions. 

\section{concluding Remarks}
In this chapter, we have introduced phase-aware Bayesian STSA estimator where the cost function includes both a power law and a weighting factor. The gain function parameters of the corresponding estimator have been chosen according to characteristics of the human auditory system. It is found that doing so suggests a decrease in the gain at high frequencies which limits the speech distortions at low frequencies while increasing the noise reduction at high frequencies.
By giving emphasis to the recent advancement in the phase-based speech processing, these amplitude estimators are derived given uncertain prior phase information. Experimental evaluation in terms of objective measures have shown that the proposed estimator can achieve substantial improvement in performance than the traditional phase-blind approaches as well as the other existing phase-aware Bayesian estimators in both oracle and blind experimental setup.
\bibliographystyle{IEEEtran}

\bibliography{ref_thesis}
\end{document}